\documentclass[acmsmall,screen,authorversion]{acmart}
\makeatletter
\AtBeginDocument{%
  \fancypagestyle{standardpagestyle}{%
    \fancyhf{}%
    \renewcommand{\headrulewidth}{\z@}%
    \renewcommand{\footrulewidth}{\z@}%
    \fancyhead[LE]{\@headfootfont\thepage}%
    \fancyhead[RO]{\@headfootfont\thepage}%
    \fancyhead[RE]{\@headfootfont\@shortauthors}%
    \fancyhead[LO]{\@headfootfont\shorttitle}%
  }%
  \fancypagestyle{firstpagestyle}{%
    \fancyhf{}%
    \renewcommand{\headrulewidth}{\z@}%
    \renewcommand{\footrulewidth}{\z@}%
  }%
  \pagestyle{standardpagestyle}%
}
\makeatother

\usepackage{multirow}
\usepackage{subcaption}
\usepackage{hyperref}
\usepackage{fancyhdr} 
\usepackage{makecell}

\definecolor{mygreen}{RGB}{35, 83, 71}
\definecolor{myred}{RGB}{142, 15, 25}
\AtBeginDocument{%
  }

\setcopyright{cc}
\setcctype{by}
\acmJournal{TORS}
\acmYear{2026} \acmVolume{1} \acmNumber{1} \acmArticle{}
\acmMonth{1} \acmDOI{10.1145/3808222}
\makeatletter
\def\@adddotafter#1{%
  \ifx\relax#1\relax
    #1%
  \else
    \def\@tempb{\expandafter\@gobble\string#1.}%
    \ifx\@tempb\@empty
      #1\@addpunct{.}%
    \else
      #1%
    \fi
  \fi
}

\begin{document}

\title{Adopting State-of-the-Art Pretrained Audio Representations for Music Recommender Systems}

% \affiliation{%

\author{Yan-Martin Tamm}
\email{yanmart.tamm@gmail.com}
\orcid{0000-0002-6174-7736}
\affiliation{%
  \institution{University of Tartu}
  \streetaddress{Narva mnt 18}
  \city{Tartu}
  \country{Estonia}
  \postcode{51009}
}

\author{Anna Aljanaki}
\email{aljanaki@gmail.com}
\orcid{0000-0002-7119-8312}
\affiliation{%
  \institution{University of Tartu}
  \streetaddress{Narva mnt 18}
  \city{Tartu}
  \country{Estonia}
  \postcode{51009}
}


% \renewcommand{\shortauthors}{Tamm and Aljanaki}

\begin{abstract}
  Over the years, Music Information Retrieval (MIR) research community has released various models pretrained on large amounts of music data. Transfer learning showcases the proven effectiveness of pretrained backend models for a broad spectrum of downstream tasks, including auto-tagging and genre classification. However, MIR papers generally do not explore the efficiency of pretrained models for Music Recommender Systems (MRS). In addition, the Recommender Systems community tends to favour traditional end-to-end neural network training. Our research addresses this gap and evaluates the performance of nine pretrained backend models (MusicFM, \mbox{Music2Vec}, MERT, EncodecMAE, Jukebox, MusiCNN, MULE, MuQ and \mbox{MuQ-MuLan}) in the context of MRS. We assess them using five recommendation approaches: K-Nearest Neighbours (KNN), Shallow Neural Network, Contrastive Multi-Modal projection, a Hybrid model, and \mbox{BERT4Rec} both for the hot and cold-start scenarios. Our findings suggest that pretrained audio representations exhibit significant performance disparity between traditional MIR tasks and both hot and cold music recommendations, indicating that valuable aspects of musical information captured by backend models may differ depending on the task. This study establishes a foundation for further exploration of pretrained audio representations to enhance music recommendation systems. 
\end{abstract}




% \received{20 February 20077}

\hyphenation{MusicFM MERT EncodecMAE Jukebox MusiCNN HitRate}

\maketitle

\section{Introduction}

Music Recommender Systems (MRS) are naturally suitable for a hybrid recommendation setting because both collaborative interactions and audio data are equally important, and usually readily available, allowing us to gain deeper insight into user preferences. Apart from performance improvement, hybrid recommender systems also have the potential to address the cold start problem for new items, which is an acute problem for music streaming platforms due to huge catalogue size, short duration of consumption and a long tail of less popular content \cite{Schedl2022}.

In content-aware MRS, Convolutional Neural Networks (CNN) are commonly trained on Mel-Spectrograms extracted from music segments, as popularized by \cite{Oord2013DeepCM}. This approach proved highly effective in processing audio and incorporating content information into recommender systems.

The Music Information Retrieval (MIR) community has introduced numerous backend models that are pretrained on extensive amounts of music data. Some of these models, such as musiCNN ~\cite{Pons2019MusiCNNPC}, are trained in a supervised manner, usually on auto-tagging task, while others, like Jukebox ~\cite{Dhariwal2020JukeboxAG}, are self-supervised. Regardless of the approach, the crucial factor is that these models can be effectively utilized for downstream tasks through transfer learning, yielding results comparable to state-of-the-art models specifically designed for those tasks. One way to do this is to fine-tune all model weights for the target task and dataset, which is computationally costly. The other way is to take the embeddings produced by a backend model and train a separate small network that adapts them to downstream tasks. We refer to these frozen embeddings as pretrained audio representations (PAR). This approach opens up the possibility of using large quantities of unlabeled music data to address problems with limited labelled examples, which could be particularly beneficial in MRS research, where access to large datasets containing both music data and user play history is limited due to copyright.

However, pretrained audio representations were not explored within the context of MRS both in MIR and RS scientific communities. The MIR community tends to focus on established tasks and benchmarks that do not include recommendations, while the Recommender Systems (RS) community tends to lean towards traditional end-to-end neural network training over these models. One notable exception is \cite{Park2022ExploitingNP}, where the authors utilized three pretrained encoders: CLMR ~\cite{Spijkervet2021ContrastiveLO}, MEE~\cite{Koo2022EndToEndMR}, and Jukebox ~\cite{Dhariwal2020JukeboxAG}. However, the primary focus of that paper was to study the role of negative preferences in user music tastes, and the use of different pretrained models emphasized the stability of the proposed method rather than being integral to the research.

Our paper compares the performance of pretrained embeddings in the context of MRS using nine recently released pretrained backend models. It should be noted that we leave the problem of finding the best architecture for combining content and collaborative information for future research. More specifically, we do not propose a new recommendation model, instead we conduct a wide comparison of pretrained representations. Our goal is to highlight the most promising ones to inspire their usage in MRS research. We outline the following research questions:
\begin{itemize}
    \item RQ1: How do different recommendation approaches compare when initialized with pretrained audio representations?
    \item RQ2: How do different pretrained audio representations compare in the context of MRS?
    \item RQ3: How does the performance of pretrained audio representations in MRS correspond to performance in MIR tasks?
\end{itemize}

In this paper, we extend and improve our previous work on the topic \cite{Tamm_2024}, adding more recommendation models, additional PARs, and results for the cold-start scenario. 

The rest of the paper is organized as follows: first, we lay out the approaches used in MRS to incorporate music data into recommendations to explain background for the paper. Next, we describe the experiment and dataset. We continue by introducing recent pretrained models from MIR that will be used to compute audio representations and briefly describe each of them. We describe the recommendation models we used, outline training details\footnote{The code is available at \href{https://github.com/Darel13712/adopting-pretrained-audio}{github.com/Darel13712/adopting-pretrained-audio}}, and conclude with results and discussion.

\section{Background}
\label{mrs}
To start our discussion, we turn our attention to different ways to incorporate audio in MRS that have been used throughout the years. To this end, we take recent reviews on the topic \cite{Deldjoo2021ContentdrivenMR,Schedl2019DeepLI,Moysis2023MusicDL} covering papers from 2006 to 2022, select all papers that use audio content and study the way it is processed. This funnel produced a total of 45~papers; the full list can be found in the appendix in \mbox{Table \ref{tab:features_references}}. It is important to note that our overview is only as comprehensive as the reviews we selected, and we can expect some relevant papers to be missing, but we believe it to be a reasonable estimation of the global trend. 

\begin{figure}
    \centering
    \includegraphics[width=1\linewidth]{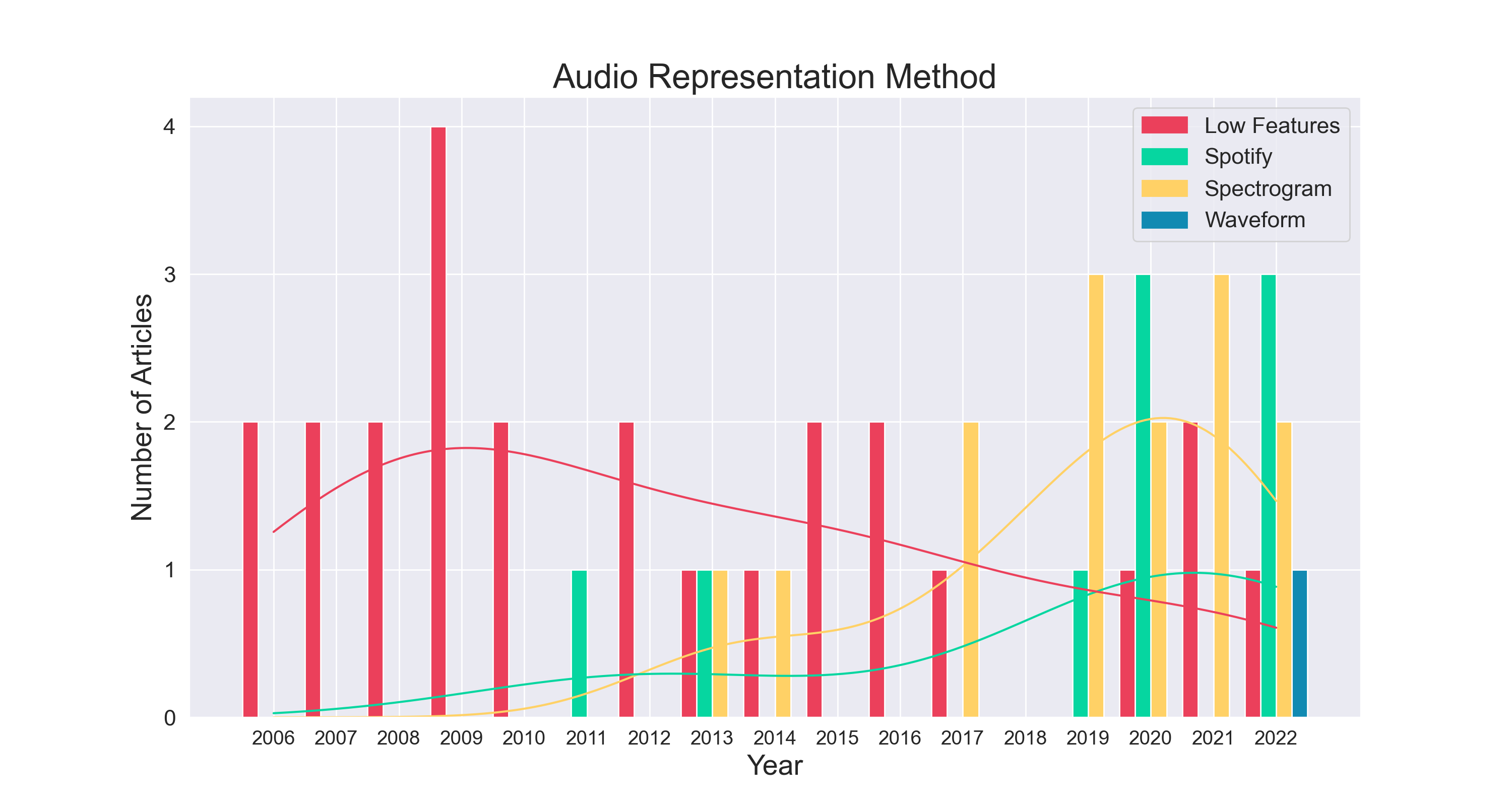}
    \caption{Number of MRS papers using different types of input data to represent audio files per year.}
    \label{fig:papers}
\end{figure}

We group the models into four distinct approaches that use different methods to represent musical data: \textbf{Low-level Features}, \textbf{Spotify Features}, \textbf{Spectrograms} and raw \textbf{Waveform}. We proceed with the description of each group. The distribution of the methods can be seen in Figure \ref{fig:papers}. \textbf{Low-level Features} include handcrafted features such as MFCC, Chroma, ZCR; \textbf{Spotify} refers to pre-computed audio descriptors that were available through Spotify API such as danceability and acousticness, \textbf{Spectrogram} refers to image input such as Mel-Scale spectrogram or Constant-Q transform, \textbf{Waveform} refers to using raw audio (time-series of amplitude samples) as an input.

\subsection{Audio features and representations}
\subsubsection{Low-level Acoustic Features}~\newline
These features originate from speech processing and MIR and involve calculating manually crafted statistics and transformations over the input signal to describe its content. We include spectral features and musical characteristics in this category. Spectral features are local short-time descriptors such as Spectral Centroid / Rolloff / Flux, Daubechies Wavelet Coefficient Histograms and the most widespread timbre descriptor — Mel Frequency Cepstral Coefficients (MFCCs). Musical characteristics include global song descriptions, such as tonality and tempo. For a more detailed description, please refer to \cite{Shao2009MusicRB,MartnGutirrez2020AME}. 

Most features like MFCCs are calculated using small time frames about 20-200ms and should be aggregated somehow to describe longer audio tracks. Some possibilities include global mean and variance aggregation \cite{Yoshii2009ContinuousPA,Yakura2022AnAS}, histogram representation \cite{McFee2011LearningCS} and bag-of-words representation \cite{Yoshii2006HybridCA}.

Low-level spectral features have been proven useful for speech recognition \cite{Padmanabhan04072015} and other tasks \cite{Li2011}. However, these features are manually crafted, which diverges from the modern end-to-end deep learning approach, which shows that useful features can be learned automatically and demonstrate better performance given enough data \cite{bitterLesson}. Moreover, there is a semantic gap between statistical descriptions of the signal, such as Zero Crossing Rate, and high semantic concepts that people use to describe music, such as genre and mood \cite{Friberg2011}.

However, multiple sources of information generally improve model performance \cite{Kim2018OneDM}, and some papers use these low-level features as an addition to other features and representations \cite{MartnGutirrez2020AME,SheikhFathollahi2021MusicSM} rather than as a standalone input, as these were used historically.
~\newline
\subsubsection{Spotify Features}~\newline
Previously known as Echo Nest features, then available through Spotify API \cite{spotify} and recently deprecated \cite{spotifyDeprecated}, these audio features consist of a mixture of low-level and mid-level music descriptors, including handcrafted features like acousticness, danceability, energy, instrumentalness, liveness and speechiness. For a detailed description, please refer to \cite{spotify,MartnGutirrez2020AME}.

Made popular with the inclusion into Million Song Dataset \cite{BertinMahieux2011TheMS}, these features became a useful and convenient way to describe audio content without sharing copyrighted audio files while significantly reducing storage consumption and compute time. They were used for a wide range of tasks, including playlist generation \cite{McFee2011TheNL}, context-aware music recommendation \cite{Chen2013UsingEC,lvarez2020MobileMR,Zangerle2020UserMF}, sequential music recommendation \cite{Pereira2019OnlineLT,Chaves2022EfficientOL}, music popularity prediction \cite{MartnGutirrez2020AME}, diversification \cite{Tommasel2022HaventIJ} and explanation \cite{Martijn2022KnowingMK}.

However, their downside is that the exact formulas for these features are not disclosed and thus present a reproducibility problem, as it is not possible to calculate these values for arbitrary audio files. Recent deprecation of these features \cite{spotifyDeprecated} highlights the dangers of dependence on proprietary APIs and stresses the need for openness and reproducibility in academia. We might expect papers using these features to fade out gradually.
~\newline
\subsubsection{Spectrogram}~\newline
Raw audio files are computationally expensive to work with and consume a lot of disk space. In contrast, Spectrogram offers a more compact and manageable alternative to a long one-dimensional audio vector. A spectrogram is a two-dimensional time-frequency matrix of amplitude values, consisting of Fourier transform spectra over time, where each spectrum shows the amplitude values by frequency at any given moment of time. In human perception, perceived distances between sounds of different frequencies are logarithmically distributed, with a better resolution at lower frequencies. Thus, in a mel-spectrogram or constant-Q transform, logarithmic transformation along the frequency axis is applied.

One of the first papers to adopt this approach was \cite{Oord2013DeepCM}, which made it popular and established as a de facto standard to represent music. In that paper, the authors train a model to predict collaborative embeddings using content information to solve the cold start problem for new items. To this end, the authors test two approaches. Firstly, they train a model using a bag-of-words representation of MFCC vectors as input. Secondly, they train a CNN with spectrograms as an input. The authors achieve better results with the latter option and conclude that predicting collaborative vectors with spectrograms is a viable approach to recommend novel music.

Following this approach, other models have been proposed. In \cite{oramas2017deep}, authors create a multi-modal model to fuse artist embeddings learned from biographies, and track embeddings learned from spectrograms for improved cold item recommendations. In \cite{Wang2014ImprovingCA}, the authors propose a method to incorporate audio into a hybrid recommendation model without the need to train a separate collaborative model beforehand. Other approaches to hybrid recommendations include \cite{Chou2017ConditionalPN,Adiyansjah2019MusicRS,Chen2020LearningAE,Gao2022AutomaticRO}. Spectrograms were also used when training Automatic Playlist Continuation \cite{Vall2019FeaturecombinationHR}, music popularity prediction models \cite{MartnGutirrez2020AME} and recommendation explanations \cite{Melchiorre2021LEMONSLE}.

Another notable approach is the use of Siamese Networks \cite{Pulis2021SiameseNN,Chen2020LearningAE} to predict similarity between tracks. Siamese Networks use two identical towers to process two tracks (and a separate tower for users), and the goal is to learn such item embeddings so that similar items have similar embeddings.

It is generally believed that Spectrogram inputs perform better than Low-level Acoustic Features \cite{Oord2013DeepCM,choi2016automatic}, given enough data. Thus, many papers increasingly choose CNNs with Spectrogram input to train item representations over using Low-level Acoustic Features. However, sometimes these approaches are combined \cite{SheikhFathollahi2021MusicSM}.

All the papers we discussed above trained the model end-to-end on a specific dataset. However, it is not the only option. In \cite{Bontempelli2022FlowMR}, authors use MusiCNN \cite{Pons2019MusiCNNPC} (an auto-tagging model) to generate item embeddings. The embeddings serve as features for six random forest binary classifiers to predict mood with an end goal of improving a recommender system. This is a relevant example demonstrating that training a model from scratch is not always necessary; instead, a pretrained model can be adopted.
~\newline
\subsubsection{Waveform}~\newline
An alternative to spectrograms is to use the raw audio waveform. Models using raw audio as an input are relatively new in MIR and other speech and sound processing fields. Compared to spectrogram-based models, their performance is not yet fully understood and varies from task to task \cite{Purwins2019DeepLF}. However, in MIR, the raw audio approach has been successfully adopted in contrast with the MRS field, where these models can rarely be found.

From all the papers we selected for this metareview, only one paper uses raw audio as an input \cite{Park2022ExploitingNP}. However, in \cite{Park2022ExploitingNP}, a pretrained backend model is not the main focus of the paper but rather a way to prove the stability of the proposed negative sampling approach. The authors used three pretrained models: CLMR \cite{Spijkervet2021ContrastiveLO}, MEE \cite{Koo2022EndToEndMR} and Jukebox \cite{Dhariwal2020JukeboxAG}. CLMR performs contrastive learning using a segment of a song as an anchor, another segment from the same song as a positive example and a segment from a different song as a negative example. The underlying encoder processes raw audio using the SampleCNN \cite{app8010150} architecture. MEE is similar to CLMR but is meant to transfer the global mastering style of one track to another. Jukebox uses Language Models for music generation and will be described further in Section \ref{sec:pretrained_models}. 
~\newline

\subsection{Summary}~\newline

In the early days of the field, MRS papers used Low-Level Acoustic Features to represent music. Since 2013, the spectrogram-based approach has become more prominent. Spotify features were used because of the relatively easy access, but became less relevant since this access was deprecated for the general public. Only one paper from our selection adopts models that operate on raw audio.

The majority of the papers employ the end-to-end learning approach. On the one hand, this happens because it is an established practice, and on the other hand, because of the scarcity of publicly available datasets containing audio files. This nudges researchers towards more readily available representations, such as MFCC and Spotify features, and towards simpler approaches that can be trained with relatively small amounts of data.

However, these problems do not hinder the training of music representation models for MIR tasks because music listening history is not a prerequisite. Only track metadata (e.g., tags, genre labels) and audio are necessary to train supervised models on extensive music collections. For self-supervised models, one only needs large amounts of unlabeled data.

This leaves an opportunity to adopt novel MIR models pretrained on large quantities of musical data for MRS. In our metareview, only two papers \cite{Bontempelli2022FlowMR,Park2022ExploitingNP} out of forty-five have used this approach. In both cases, it was more of a practically motivated solution than the focus of the research. While there were some developments \cite{pembek2025let,grotschla2024towards} since the time of our original publication \cite{Tamm_2024}, the adoption of pretrained audio representations remains relatively unexplored in music recommendations. Indeed, even modern approaches such as Semantic IDs \cite{rajput2023recommender,singh2024better} that expect pretrained embeddings are sometimes trained with basic content inputs such as Spotify features \cite{mei2025semantic} In the next section, we will bridge this gap by comparing the performance of state-of-the-art MIR models in the context of recommendations and pave the way for future research on pretrained audio representations in MRS.

\section{Experiment}

We select a set of recent state-of-the-art models and extract audio embeddings for music tracks using 30 second preview snippets of audio from the original Music4All dataset \cite{santana2020music4all}. The resulting embedding vectors we refer to as pretrained audio representations (PAR). For our research purposes, these representations are frozen, i.e., immutable. In order to assess their performance in MRS, we plug these vectors into different recommendation models, generate novel item recommendations for each user and evaluate the quality using standard metrics. Recommendation models cannot change PARs during the training, but they do learn their projections into a new latent space for recommendations. We study both the hot and cold-start scenarios. This section describes the dataset and its split, then presents the PARs and recommendation models, concluding with training details.

\subsection{Dataset}
\label{sec:data}
\begin{figure}
    \centering
    \includegraphics[width=1\linewidth]{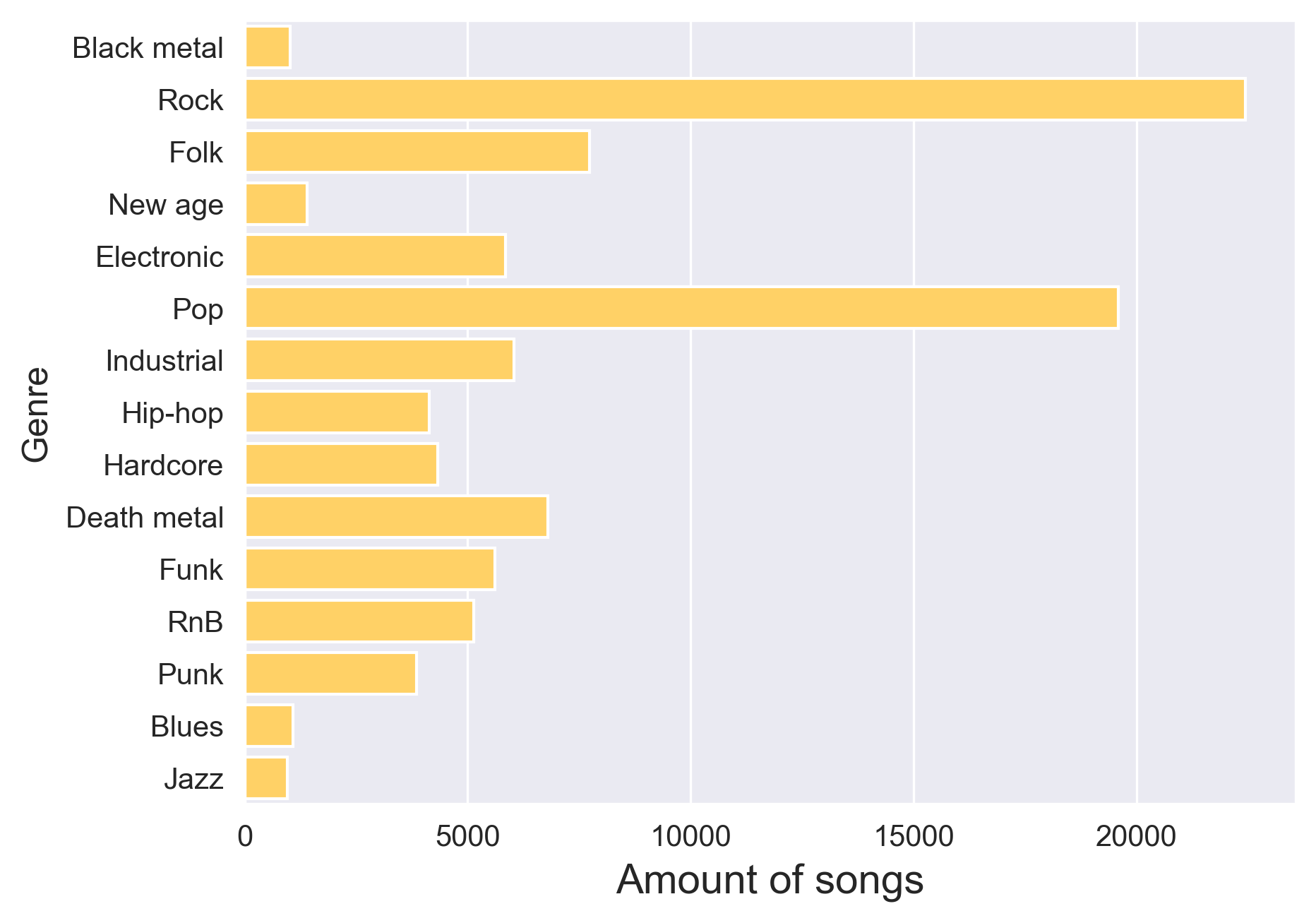}
    \caption{A count plot showing genre distribution in Music4all dataset.}
    \label{fig:genres}
\end{figure}

We use the Music4All-Onion \cite{Moscati2022Music4AllOnionA} dataset because it contains both music samples, user listening history and precomputed MFCC features. This dataset contains music in a variety of genres, dominated by rock and pop music, representative of the mainstream music that most of our backend models were trained on. Figure \ref{fig:genres}  shows the distribution of genres in Music4All dataset, as derived using hierarchical clustering of the genre co-occurrence matrix from the 853 genre tags in Music4All. We select one month of listening history (after 2019-02-01) as test and validation data and the previous year as training data. Hot test and validation contain only the items that users have not listened to before 2019-02-01, so all the items in the test and validation set are new in a user's listening history. We split last month into validation and hot test set by users with a 30/70 proportion. Cold test contains all items that first appear during the test month of 2019-02. Resulting split sizes can be found in Table \ref{tab:data}. The same data split is shared across all experiments.

It is important to describe the way we treat hot and cold test sets to evaluate recommendations. In the case of the hot test, we generate recommendations using all the hot items in the dataset as candidates and filter out previously seen items. In the case of the cold test, we generate recommendations using only the 589 cold items present in the cold test. There are not so many cold items because we used the "natural" cold items and did not generate them artificially by removing them from user histories. Since we generate recommendations using only 589 items, the metrics will naturally have a higher value than those calculated for the hot set using all the hot items. We could have calculated the cold metrics using all the items as candidates, but we chose not to do this for two reasons. First, the metric values are relatively small and inconvenient when evaluating comparative performance. Second, recommending only new items that came out during one month is a plausible and realistic recommendation setting.

\begin{table}
    \centering
\caption{Train-test split for Music4All-Onion dataset. We used one month for validation and testing, leaving 70\% of users for hot test and 30\% for validation. The previous 12 months are used as train data. The cold test includes all the new items that first appeared in the test month.}
\label{tab:data}
    \begin{tabular}{cccccl}
         &  Train&  Validation& Hot Test & Cold Test\\
         Num Users&  21,018&  2,459& 10,238 & 4,553\\
         Num Items&  55,262&  23,625& 45,002 & 589\\
 Num Interactions& 5,371,381& 48,787&280,745 & 23,269\\
    \end{tabular}

\end{table}

\subsection{Pretrained Audio Representations}
\label{sec:pretrained_models}

The list of the models we used can be found in Table \ref{tab:models}. Below we describe each of them.

\textfloatsep=2.5ex

\begin{table}[htbp]
    \centering
\caption{Embedding sizes for pretrained audio representations we used}
\label{tab:models}
    \begin{tabular}{cc}
         Model & Embedding Size \\
 MFCC&104\\
         MusiCNN \cite{Pons2019MusiCNNPC}&  200\\
 MuQ-MuLan \cite{zhu2025muq}&512\\
         EncodecMAE \cite{Pepino2023EnCodecMAELN}&  768\\
         Music2Vec \cite{Li2022MAPMusic2VecAS}&  768\\
         MusicFM \cite{Won2023AFM}&  1024\\
         MERT \cite{Li2023MERTAM}&  1024\\
 MuQ \cite{zhu2025muq}&1024\\
 MULE \cite{mccallum2022supervisedunsupervisedlearningaudio}&1728\\
         Jukebox \cite{Dhariwal2020JukeboxAG, Castellon2021CodifiedAL}&  4800\\
    \end{tabular}
\end{table}

Mel Frequency Cepstral Coefficients (MFCCs) represent a short-term power spectrum of a sound based on a linear cosine transform of a log power spectrum on a nonlinear mel scale of frequency. MFCC is a low-level acoustic feature designed to capture the timbral characteristics of an audio signal. It is widely utilized in various domains \cite{Abdul2022MelFC}, including RS. Strictly speaking, MFCC is not a \textit{pretrained} audio representation since there is no backend model with learnable parameters behind it. However, we employ this \textit{precalculated} audio representation as a baseline and a reference point due to its widespread popularity. We utilize the mean and flattened covariance matrix of MFCCs provided with the dataset we have utilized.

MusiCNN \cite{Pons2019MusiCNNPC} is a CNN trained in a supervised way to predict crowd-sourced labels (50 classes: tags from last.fm). It takes log mel spectrograms of audio files as an input and applies a series of convolutional and dense layers to them. The feature representation is extracted from the penultimate layer of the network. It was trained on 200k audio files from the Million Song Dataset ~\cite{BertinMahieux2011TheMS}.

Jukebox \cite{Dhariwal2020JukeboxAG} is a music generation model that consists of three separate Vector Quantized Variational Autoencoders (VQ-VAE) with different temporal resolutions. The encoder part of VQ-VAE compresses raw audio input into a sequence of embeddings using \mbox{1-D convolutions}. This sequence is then turned into a sequence of discrete tokens using codebooks. The decoder reconstructs raw audio from latent representations. An autoregressive Sparse Transformer then processes sequences of compressed tokens to learn the prior to generate further samples. This approach of training a Language Model over tokenized music representation has been shown to be a robust foundation for downstream MIR tasks \cite{Castellon2021CodifiedAL}.

Music2vec \cite{Li2022MAPMusic2VecAS} uses a multi-layer 1-D CNN feature extractor to compress 16kHz audio input into 50Hz representations that are fed into 12-layer Transformer Blocks. The student model takes partially masked input and tries to predict the average of the top-K layers of the teacher model. This approach enables impressive performance while utilizing a model that is only about ~2\% of Jukebox's in parameter size. 

Encodec \cite{Defossez2022HighFN} is a neural audio codec that compresses raw audio from 24kHz to 75Hz. It is done by applying a series of convolutional and LSTM blocks to get a sequence of 128-dimensional vectors, which are processed with a residual vector quantization block (RVQ) that maps the input vector to the index of one of the 1024 closest codebook words, then calculates the residual and maps it to a second codebook and so on for a total of 32 codebooks. EncodecMAE ~\cite{Pepino2023EnCodecMAELN} further adopts these representations and compresses them into a single embedding. That is done by applying a Masked Auto Encoder (MAE) on raw Encodec outputs before the RVQ block to predict discrete targets from the RVQ codebooks.

MERT \cite{Li2023MERTAM} is trained in a masked language modeling paradigm, incorporating teacher models to generate pseudo labels:  an acoustic teacher based on Residual Vector Quantisation — Variational AutoEncoder and a musical teacher based on the Constant-Q Transform. Notably, the model can scale from 95M to 330M parameters.

MusicFM \cite{Won2023AFM} is an improvement over MERT design where a BERT-style encoder~\cite{Hsu2021HuBERTSS} is replaced with a Conformer~\cite{Zhang2020PushingTL} and \mbox{k-means} clustering tokenization is replaced with random projection and random codebook approach from BEST-RQ~\cite{Chiu2022SelfsupervisedLW}.

MuQ \cite{zhu2025muq} is a self-supervised music representation learning model for music understanding. During training, it takes a spectrogram as an input, partially masks it with noise and uses Conformer~\cite{Zhang2020PushingTL} for context learning. However, instead of a raw spectrogram, the model is trained to predict tokens generated by a newly proposed Mel Residual Vector Quantization module (Mel-RVQ). Mel-RVQ takes inspiration from the residual approach of Encodec \cite{Defossez2022HighFN} but simplifies the architecture by replacing a series of convolution blocks over waveform with a linear projection over spectrogram.

MuQ-MuLan \cite{zhu2025muq} is a joint music-text embedding model. It is based on a two-tower contrastive approach. Music is processed with a MuQ model, averaged by time to get embedding for the whole track and then projected into a target space. Text is processed with a pretrained RoBERTa \cite{liu2019robertarobustlyoptimizedbert} model and similarly average-pooled and projected. These two embeddings are trained to be similar with a Decoupled Contrastive Learning (DCL) loss \cite{yeh2022decoupledcontrastivelearning}.

MULE \cite{mccallum2022supervisedunsupervisedlearningaudio} is a music understanding model. The authors aimed to compare supervised and unsupervised learning strategies for pretraining music understanding models for several downstream tasks. To this end, authors use a version of the Short-Fast Normalizer-Free Net F0 \cite{wang2022learninguniversalaudiorepresentations} as a backbone for music processing.

In cases where models described above produce representations for small chunks of audio with lengths ranging from a couple of milliseconds to a couple of seconds, we average these embeddings over time to get a single track-level representation.

\subsection{Data distribution in the training datasets}

Our experiment relies on the assumption that content embedding models were trained on music that is similar enough to the music in Music4All. As a minimum, we assume that music comes from a similar musical tradition and era, and covers similar genres.
It has been shown that the impact of the data distribution and quality of labels on model performance can be significant, with supervised models with expert labeling outperforming the self-supervised models and the self-supervised models being particularly sensitive to data distribution disparity \cite{mccallum2022supervisedunsupervisedlearningaudio}. 

\begin{table}[htbp]
    \centering
    \caption{Training data of the models in our experiment}
    \begin{tabular}{lccc}
    \toprule
    Model & Dataset \\
    \midrule
    MuQ  &     Music4All + 160k hours in-house     \\
    MuQ-MuLan  &    Music4All + 160k hours in-house   \\
        MERT  &      160k hours scraped     \\
    EncodecMAE  & Audioset + FMA + Libri-Light \\
    MusicFM  &  MSD      \\
    MusiCNN  &     MSD     \\
    Jukebox  &     1.2 million songs      \\
    MULE   &     Musicset    \\
    Music2Vec  &   1000 hours of audio     \\

    \bottomrule
    \end{tabular}
    \label{tab:genre-distribution-by-model}
\end{table}

In Table \ref{tab:genre-distribution-by-model}, we describe to the best of available knowledge the datasets that the models were trained on. However, inspecting the dataset genre distribution is not straightforward for several reasons. Firstly, for seven out of nine models, all or most of the data was private, and no genre information was provided in the paper. Secondly, even if the dataset is known and publicly available, the genre labels are usually derived from tags with diverse, ambiguous taxonomies that do not allow straightforward comparisons (e.g., creating a bar plot of genre distributions). We thus focus on analyzing the training genres of MusicFM and MusiCNN because they are the only two models for which the full genre distribution of the Million Song Dataset is known \cite{Schreiber2015ImprovingGA}. The genres that constitute 4\% or more of songs in MSD are Rock, Electronic, Jazz, Pop, RnB, Rap, Metal, Country, and Punk. As compared to Music4All, two genres were missing from the taxonomy - Industrial and Funk, but they could very well be represented under their parent genres, Electronic (or Rock) and RnB. However, we emphasize that genre overlap alone does not fully characterize dataset similarity. For instance, the specifics of the FMA dataset are less about genre and more about production: the music largely comes from indie artists and sometimes includes live recordings.

MuQ and MuQ-MuLan included Music4All in their training data, though in a self-supervised setting, which does not constitute a data leak. EncodecMAE includes FMA, which leans heavily towards electronic music and Creative Commons content. In an ideal experimental design, we would retrain all models on the same data (either including or excluding Music4All). This is computationally infeasible given the unavailability of the proprietary datasets and the resources required to train these models.

\subsection{Recommendation models}
After obtaining embeddings for each audio track using pretrained backend models, we must decide how to use them to produce recommendations. It is important to note that the goal of our study is not finding the best possible architecture of a neural network for this task but rather finding the best performing \textbf{content models} that are most promising for the next generation of hybrid recommender systems. To this end, we use five methods of varying complexity:

\begin{itemize}
    \item K-Nearest Neighbours (KNN)
    \item Shallow neural network
    \item Bimodal contrastive network
    \item Hybrid model
    \item BERT4Rec \cite{Sun2019BERT4RecSR}
\end{itemize}
~\newline
\subsubsection{KNN}~\newline
To implement KNN, we create a representation of a user by averaging embeddings of the items in their profile and then generate recommendations by retrieving nearest neighbours of their average embedding that are not part of a user profile. It is important to highlight that we do not train any model here but rather use PARs as-is. While raw content embeddings are unlikely to carry a lot of useful preference information suitable for recommendations, it allows us to see whether higher initial position in ratings of a PAR translates to improved performance when coupled with a collaborative signal in other models.

We also use the KNN to generate recommendations for all the other models except for the BERT4Rec. Namely, we use collaborative information to train a model that projects frozen PARs to a new space and then generate recommendations using the KNN method described above.
\\

\subsubsection{Shallow Net}~\newline
We incorporate the listening history into the recommendation model for the following approach. Specifically, we process user and item IDs with an embedding layer and a fully connected layer preserving dimensions with a ReLU activation. The score for the user-item pair is the cosine between the resulting vectors. The item embedding layer is initialized with pretrained embeddings from the corresponding backend model. Moreover, we freeze the weights for the item embedding layer to preserve useful content information stored in them. The user embedding layer is initialized with a mean of the user's tracks, but the weights are unfrozen and can be changed. The model is trained with Max Margin Hinge Loss as in~\cite{Lee2018DeepCE} and negative sampling strategy from~\cite{Magron2021NeuralCC}. More specifically, for each user-item pair from the dataset, we consider an item and a user as positive examples and sample an additional 20 negative users who did not interact with this item for training. Our preliminary studies tested the configuration for frozen item weights, user initialization, and negative sampling strategy and showed the best results. We refer to this model as Shallow Net.\\

\subsubsection{Bimodal}~\newline
Following the idea that multi-modal embeddings trained with contrastive loss may improve performance and stability \cite{ferraro2023contrastivelearningcrossmodalartist,zhu2025muq,mmsbenfricsamms24} we employ a similar approach. To this end, we train a state-of-the-art collaborative model ELSA \cite{elsa} to get a collaborative embedding for every item. Then, we use a simple projection layer consisting of a Linear layer, ReLU, another Linear layer and an L2-normalization to unit norm to put both content and collaborative embeddings into the same 786-dimensional target space. Similar to \cite{zhu2025muq}, we want to make different modality projections for a single item close using contrastive learning. We take the DCL loss \cite{yeh2022decoupledcontrastivelearning} as a basis, but we simplify it into an alternative symmetrical version that we refer to as ADCL. 

So, if we write DCL loss like this:
\[
\mathcal{L}_{DCL} = \frac{1}{N} \sum_{k=1}^{2} \sum_{i=1}^{N} \left[ - \text{sim}(z_1^i, z_2^i) + \log \sum_{j=1}^{N} \left( e^{\text{sim}(z_k^i, z_k^j)} + e^{\text{sim}(z_1^i, z_2^j)} \right) \right]
\]
\[
\text{sim}(z_1^i, z_2^j) = \frac{z_1^i \cdot z_2^j}{\tau}
\]

Then, the ADCL loss we used will be defined like this:
\[
\mathcal{L}_{ADCL} = \frac{1}{N} \sum_{i=1}^{N} \left[ - \text{sim}(z_1^i, z_2^i) + \log \sum_{j=1}^{N} \left( e^{\text{sim}(z_1^i, z_2^j)} \right) \right]
\]
where $N$ is the number of items, $z_1^i$ and $z_2^i$ refer to the embedding of an item $i$ in modality $1$ and $2$ respectively and $\tau$ is a temperature parameter. In our case, the modalities are the collaborative modality from ELSA and the content modality from a PAR.

As we can see, the original DCL loss requires the calculation of the whole formula twice because of the asymmetrical term $sim(z_k^i,z_k^j)$, which is needed to make embeddings of different items dissimilar inside one modality. One could make the formula symmetrical by introducing both $sim(z_1^i, z_1^j)$ and $sim(z_2^i, z_2^j)$  into the summation together. We have also tried that, but in the end, deleting this term showed better results in our preliminary experiments, and this is why we chose ADCL over DCL.

It is an interesting direction for future research as to why this version works better and whether it is generalizable to tasks other than MRS, but this is out of the scope of this paper.\\

\subsubsection{Hybrid}~\newline
Hybrid models combining both collaborative and content information are widespread in MRS, especially as a solution for the cold-start problem. However,  there is no universally accepted way to do this. Since we already used ELSA as a part of our Bimodal network, we also decided to adopt it for the hybrid model, essentially doing something similar to the Beeformer model \cite{beeformer}. The difference would be that the Beeformer uses a textual description of an item as content information, and we use PARs.

The architecture is essentially the same as in the projection part of the Bimodal network: the model takes frozen PARs as an input and transforms them through the Linear layer, ReLU, Linear layer and a normalization layer to the target space with the dimension size~768. The resulting projected embeddings $A$ are used to calculate the ELSA loss:
\[
\min_{A} \| X - X(AA^\top - I) \|_F^2
\]
where $X$ is the interaction matrix and $I$ is the identity matrix.

~\newline
\subsubsection{BERT4Rec}~\newline
BERT4Rec \cite{Sun2019BERT4RecSR} is a popular and effective model for sequential recommendations that leverages the Bidirectional Encoder Representations from Transformers (BERT). The learning objective of \mbox{BERT4Rec} is to mask certain elements within a sequence and predict them based on their surrounding context. We chose it to estimate the potential of a more complex and drastically different architecture in our setting. We incorporate pretrained embeddings as a frozen projection of the embedding module, just as we did with the Shallow Net. Due to memory constraints, we limited the maximum length of a sequence to 300, thus limiting the amount of data used for prediction. This value approximately equals the mean length of a user history in our dataset, so it fully covers more than half of the profiles. However, the results may still be further improved by increasing this parameter to cover the complete profiles of all users. To generate recommendations we ranked all items according to scores predicted at the last timestamp of a user profile.

\subsection{Training Details}
We use Adam optimizer with lr=0.001, early stopping, and we reduce the learning rate on the plateau. The learning rate for the Hybrid model is lr=0.01. For Shallow Net, we train for 100 epochs with cosine-based Hinge Loss. For each positive user-item pair, we sampled 20 negative users at random for this item. For \mbox{BERT4Rec}, we train for 200 epochs with Cross Entropy loss. The bimodal net is trained for 100 epochs using the ADCL loss, and the hybrid model is trained for 200 epochs using ELSA loss. We calculate HitRate@50, Recall@50 and binary NDCG@50 to evaluate the results for the hot test using standard definitions \cite{Tamm_2021}. For the cold test, we lowered the length of a recommendation list to 20 since there are fewer items in the cold test set. We also computed MRR and Precision, but we omit them from this paper because they are highly correlated with the other metrics and do not give a broader perspective.

We tuned model configurations and hyperparameters such as learning rate, sequence length for BERT4Rec and loss for the Bimodal Net using validation set performance during a set of preliminary experiments. Number of epochs was chosen on the basis of model convergence on validation set. Usually early stopping engages before the full number of epochs. While we only report the results with frozen item embeddings, we also experimented with unfrozen weights, but did not get better performance compared to the frozen variant, so we omit this option from our report.

\section{Results}
\subsection{Hot Test}

We start by looking at Table \ref{tab:knn} to evaluate the performance of raw embeddings with KNN to generate recommendations. We can see all the models rank higher than the MFCC baseline. The three top-ranking models for KNN are MusicFM, EncodecMAE and MuQ-MuLan. We should also note a large gap of metric values between the KNN model and other models (Table \ref{tab:performance}). With the KNN, models score between 0.01 and 0.03 HitRate. With other recommendation models, the values are generally ten times larger, which is expected because raw content embeddings are unlikely to carry a lot of preference information. This highlights the importance of combining the content and collaborative information in order to improve performance in recommendations. Taking this into account we proceed with results for the rest of the models.

\begin{table}[htbp]
    \centering
    \caption{Performance of pretrained embeddings using KNN}
    \begin{tabular}{lcccccc}
    \toprule
    Embeddings & HitRate@50 & Recall@50 & NDCG@50 \\
    \midrule
    MFCC & 0.0147   &  0.0005 &  0.0004 \\
    MuQ &  0.0169   &  0.0008 &  0.0005 \\
    MERT & 0.0171   &  0.0007 &  0.0005 \\
    Jukebox & 0.0195   &  0.0007 &  0.0005 \\
    Music2Vec & 0.0218   &  0.0007 &  0.0005 \\
    MusiCNN & 0.0275   &  0.0009 &  0.0008 \\
    MULE & 0.0282  &  0.0011 &  0.0009 \\
    MuQ-MuLan & 0.0289 &  0.0011 & 0.0009 \\
    EncodecMAE & 0.0317   &  0.0010 &  0.0009 \\
    MusicFM &\textbf{ 0.0370 }  & \textbf{ 0.0012} &  \textbf{0.0010} \\
    \bottomrule
    \end{tabular}
    \label{tab:knn}
\end{table}
\subsubsection{PAR performance}~\newline

\begin{table}[htbp]
    \centering
    \caption{Comparison of performance of pretrained embeddings using different recommender models. Cell format: metric value ± margin of error (\% difference with a shuffled initialization). Bold indicates the highest mean. Underlined values indicate methods whose 95\% confidence interval overlaps with that of the best-performing method. Wide confidence intervals are highlighted in red.}
    \begin{tabular}{cllllccc}
    \toprule
    & \multicolumn{1}{l}{Embeddings} & \multicolumn{1}{c}{HitRate@50} & \multicolumn{1}{c}{Recall@50} & \multicolumn{1}{c}{NDCG@50} \\

    \midrule
    \multirow{10}{*}{\rotatebox{90}{Shallow Net}}
 & MFCC & 0.2483 \footnotesize \textcolor{gray}{ ± 0.0090 \textcolor{mygreen}{(↑112\%)}}  &  0.0179 \footnotesize \textcolor{gray}{ ± 0.0010 \textcolor{mygreen}{(↑220\%)}}  &  0.0139 \footnotesize \textcolor{gray}{ ± 0.0008 \textcolor{mygreen}{(↑196\%)}} \\
 & Jukebox & 0.2982 \footnotesize \textcolor{gray}{ ± 0.0227 \textcolor{mygreen}{(↑653\%)}}  &  0.0266 \footnotesize \textcolor{gray}{ ± 0.0029 \textcolor{mygreen}{(↑1673\%)}}  &  0.0197 \footnotesize \textcolor{gray}{ ± 0.0022 \textcolor{mygreen}{(↑1542\%)}} \\
 & Music2Vec & 0.3200 \footnotesize \textcolor{gray}{ ± 0.0057 \textcolor{mygreen}{(↑331\%)}}  &  0.0287 \footnotesize \textcolor{gray}{ ± 0.0009 \textcolor{mygreen}{(↑676\%)}}  &  0.0216 \footnotesize \textcolor{gray}{ ± 0.0004 \textcolor{mygreen}{(↑620\%)}} \\
  & MusicFM & 0.3217 \footnotesize \textcolor{gray}{ ± 0.0174 \textcolor{mygreen}{(↑442\%)}}  &  0.0291 \footnotesize \textcolor{gray}{± 0.0025 \textcolor{mygreen}{(↑978\%)}}  &  0.0215 \footnotesize \textcolor{gray}{ ± 0.0018 \textcolor{mygreen}{(↑877\%)}} \\
 & MERT & 0.3264 \footnotesize \textcolor{gray}{ ± 0.0255 \textcolor{mygreen}{(↑716\%)}}  &  0.0304 \footnotesize \textcolor{gray}{ ± 0.0037 \textcolor{mygreen}{(↑1800\%)}}  &  0.0222 \footnotesize \textcolor{gray}{ ± 0.0027 \textcolor{mygreen}{(↑1608\%)}} \\
 & EncodecMAE & 0.3319 \footnotesize \textcolor{gray}{ ± 0.0116 \textcolor{mygreen}{(↑262\%)}}  &  0.0306 \footnotesize \textcolor{gray}{ ± 0.0015 \textcolor{mygreen}{(↑512\%)}}  &  0.0226 \footnotesize \textcolor{gray}{ ± 0.0009 \textcolor{mygreen}{(↑479\%)}} \\
 & MusiCNN & 0.3581 \footnotesize \textcolor{gray}{ ± 0.0068 \textcolor{mygreen}{(↑113\%)}}  &  0.0344 \footnotesize \textcolor{gray}{ ± 0.0010 \textcolor{mygreen}{(↑251\%)}}  &  0.0261 \footnotesize \textcolor{gray}{ ± 0.0009 \textcolor{mygreen}{(↑222\%)}} \\
 & MuQ-MuLan & \underline{0.3730} \footnotesize \textcolor{gray}{ ± 0.0079 \textcolor{mygreen}{(↑250\%)}}  &  \underline{0.0377} \footnotesize \textcolor{gray}{ ± 0.0017 \textcolor{mygreen}{(↑539\%)}}  &  0.0279 \footnotesize \textcolor{gray}{ ± 0.0012 \textcolor{mygreen}{(↑494\%)}} \\
 & MULE & 0.3735 \footnotesize \textcolor{gray}{ ± 0.0065 \textcolor{mygreen}{(↑323\%)}}  &  0.0388 \footnotesize \textcolor{gray}{ ± 0.0002 \textcolor{mygreen}{(↑824\%)}}  &  0.0284 \footnotesize \textcolor{gray}{ ± 0.0007 \textcolor{mygreen}{(↑761\%)}} \\
 & MuQ & \textbf{0.3895} \footnotesize \textcolor{gray}{ ± 0.0089 \textcolor{mygreen}{(↑189\%)}}  &  \textbf{0.0412} \footnotesize \textcolor{gray}{ ± 0.0020 \textcolor{mygreen}{(↑422\%)}}  &  \textbf{0.0308} \footnotesize \textcolor{gray}{ ± 0.0012 \textcolor{mygreen}{(↑397\%)}} \\

    \midrule
    \addlinespace[0.7em]
    \midrule

    \multirow{10}{*}{\rotatebox{90}{Bimodal Average}} & MFCC & 0.0435 \footnotesize \textcolor{gray}{ ± 0.0054 \textcolor{mygreen}{(↑80\%)}}  &  0.0021 \footnotesize \textcolor{gray}{ ± 0.0002 \textcolor{mygreen}{(↑110\%)}}  &  0.0014 \footnotesize \textcolor{gray}{ ± 0.0002 \textcolor{mygreen}{(↑100\%)}} \\
 & Jukebox & 0.1157 \footnotesize \textcolor{gray}{ ± 0.0464 \textcolor{mygreen}{(↑361\%)}}  &  0.0085 \footnotesize \textcolor{gray}{ ± 0.0033 \textcolor{mygreen}{(↑750\%)}}  &  0.0051 \footnotesize \textcolor{gray}{ ± 0.0023 \textcolor{mygreen}{(↑538\%)}} \\
 & MusiCNN & 0.1657 \footnotesize \textcolor{gray}{ ± 0.0086 \textcolor{mygreen}{(↑511\%)}}  &  0.0098 \footnotesize \textcolor{gray}{ ± 0.0007 \textcolor{mygreen}{(↑880\%)}}  &  0.0070 \footnotesize \textcolor{gray}{ ± 0.0006 \textcolor{mygreen}{(↑775\%)}} \\
 & MusicFM & 0.1698 \footnotesize \textcolor{gray}{  ± 0.0353 \textcolor{mygreen}{(↑541\%) }}  &  0.0124  \footnotesize \textcolor{gray}{ ± 0.0027 \textcolor{mygreen}{(↑1278\%)}}  &  0.0083 \footnotesize \textcolor{gray}{ ± 0.0017 \textcolor{mygreen}{(↑822\%)}} \\
 & MULE & 0.2271 \footnotesize \textcolor{gray}{ ± 0.0248 \textcolor{mygreen}{(↑691\%)}}  &  0.0172 \footnotesize \textcolor{gray}{ ± 0.0023 \textcolor{mygreen}{(↑1333\%)}}  &  0.0121 \footnotesize \textcolor{gray}{ ± 0.0019 \textcolor{mygreen}{(↑1244\%)}} \\
 & MERT & 0.2462 \footnotesize \textcolor{gray}{ ± 0.0481 \textcolor{mygreen}{(↑851\%)}}  &  0.0201 \footnotesize \textcolor{gray}{ ± 0.0045 \textcolor{mygreen}{(↑1910\%)}}  &  0.0146 \footnotesize \textcolor{gray}{ ± 0.0033 \textcolor{mygreen}{(↑1522\%)}} \\
 & EncodecMAE & 0.2551 \footnotesize \textcolor{gray}{ ± 0.0111 \textcolor{mygreen}{(↑855\%)}}  &  0.0202 \footnotesize \textcolor{gray}{ ± 0.0017 \textcolor{mygreen}{(↑1736\%)}}  &  0.0144 \footnotesize \textcolor{gray}{ ± 0.0013 \textcolor{mygreen}{(↑1500\%)}} \\
 & Music2Vec & 0.2850 \footnotesize \textcolor{gray}{ ± 0.0168 \textcolor{mygreen}{(↑652\%)}}  &  0.0211 \footnotesize \textcolor{gray}{ ± 0.0016 \textcolor{mygreen}{(↑1141\%)}}  &  0.0161 \footnotesize \textcolor{gray}{ ± 0.0009 \textcolor{mygreen}{(↑1138\%)}} \\
 & MuQ & \underline{0.3794} \footnotesize \textcolor{gray}{ ± 0.0115 \textcolor{mygreen}{(↑1418\%)}}  &  \underline{0.0355} \footnotesize \textcolor{gray}{ ± 0.0017 \textcolor{mygreen}{(↑3844\%)}}  &  \underline{0.0269} \footnotesize \textcolor{gray}{ ± 0.0013 \textcolor{mygreen}{(↑3262\%)}} \\
 & MuQ-MuLan & \textbf{0.3975} \footnotesize \textcolor{gray}{ ± 0.0173 \textcolor{mygreen}{(↑1029\%)}}  &  \textbf{0.0358} \footnotesize \textcolor{gray}{ ± 0.0023 \textcolor{mygreen}{(↑2287\%)}}  &  \textbf{0.0276} \footnotesize \textcolor{gray}{ ± 0.0017 \textcolor{mygreen}{(↑2409\%)}} \\

    \midrule
    \addlinespace[0.7em]
    \midrule

    \multirow{10}{*}{\rotatebox{90}{Hybrid}}
 & Jukebox & 0.0775 \footnotesize \textcolor{gray}{ ± 0.0655 \textcolor{mygreen}{(↑219\%)}}  &  0.0035 \footnotesize \textcolor{gray}{ ± 0.0032 \textcolor{mygreen}{(↑289\%)}}  &  0.0027 \footnotesize \textcolor{gray}{ ± 0.0024 \textcolor{mygreen}{(↑286\%)}} \\
 & MFCC & 0.3021 \footnotesize \textcolor{gray}{ ± 0.0216 \textcolor{mygreen}{(↑219\%)}}  &  0.0208 \footnotesize \textcolor{gray}{ ± 0.0028 \textcolor{mygreen}{(↑384\%)}}  &  0.0169 \footnotesize \textcolor{gray}{ ± 0.0021 \textcolor{mygreen}{(↑357\%)}} \\
 & Music2Vec & 0.3025 \footnotesize \textcolor{myred}{ ± 0.2874 \textcolor{mygreen}{(↑915\%)}}  &  0.0251 \footnotesize \textcolor{myred}{ ± 0.0256 \textcolor{mygreen}{(↑1992\%)}}  &  0.0209 \footnotesize \textcolor{myred}{ ± 0.0213 \textcolor{mygreen}{(↑2512\%)}} \\
  & MusicFM & 0.3067 \footnotesize \textcolor{gray}{ ± 0.0278  \textcolor{mygreen}{(↑278\%)}}  &  0.0218  \footnotesize \textcolor{gray}{± 0.0036 \textcolor{mygreen}{(↑459\%) }}  &  0.0173 \footnotesize \textcolor{gray}{ ± 0.0028 \textcolor{mygreen}{(↑394\%)}} \\
 & MERT & 0.3680 \footnotesize \textcolor{gray}{ ± 0.0362 \textcolor{mygreen}{(↑116\%)}}  &  0.0308 \footnotesize \textcolor{gray}{ ± 0.0056 \textcolor{mygreen}{(↑238\%)}}  &  0.0250 \footnotesize \textcolor{gray}{ ± 0.0048 \textcolor{mygreen}{(↑201\%)}} \\
 & EncodecMAE & 0.3804 \footnotesize \textcolor{gray}{ ± 0.0487 \textcolor{mygreen}{(↑137\%)}}  &  0.0334 \footnotesize \textcolor{gray}{ ± 0.0090 \textcolor{mygreen}{(↑275\%)}}  &  0.0270 \footnotesize \textcolor{gray}{ ± 0.0077 \textcolor{mygreen}{(↑246\%)}} \\
 & MULE & 0.4071 \footnotesize \textcolor{gray}{ ± 0.0063 \textcolor{mygreen}{(↑13\%)}}  &  \underline{0.0441} \footnotesize \textcolor{gray}{ ± 0.0012 \textcolor{mygreen}{(↑52\%)}}  &  0.0349 \footnotesize \textcolor{gray}{ ± 0.0010 \textcolor{mygreen}{(↑45\%)}} \\
 & MuQ & 0.4239 \footnotesize \textcolor{gray}{ ± 0.0036 \textcolor{mygreen}{(↑535\%)}}  &  \underline{0.0468} \footnotesize \textcolor{gray}{ ± 0.0007 \textcolor{mygreen}{(↑1460\%)}}  &  \underline{0.0370} \footnotesize \textcolor{gray}{ ± 0.0006 \textcolor{mygreen}{(↑1380\%)}} \\
 & MusiCNN & \underline{0.4323} \footnotesize \textcolor{gray}{ ± 0.0074 \textcolor{mygreen}{(↑9\%)}}  &  \underline{0.0451} \footnotesize \textcolor{gray}{ ± 0.0015 \textcolor{mygreen}{(↑19\%)}}  &  \underline{0.0368} \footnotesize \textcolor{gray}{ ± 0.0010 \textcolor{mygreen}{(↑16\%)}} \\
 & MuQ-MuLan & \textbf{0.4450} \footnotesize \textcolor{gray}{ ± 0.0084 \textcolor{mygreen}{(↑195\%)}}  &  \textbf{0.0485} \footnotesize \textcolor{gray}{ ± 0.0034 \textcolor{mygreen}{(↑464\%)}}  &  \textbf{0.0390} \footnotesize \textcolor{gray}{ ± 0.0020 \textcolor{mygreen}{(↑413\%)}} \\

    \midrule
    \addlinespace[0.7em]
    \midrule
    
    \multirow{10}{*}{\rotatebox{90}{BERT4Rec}} & Jukebox & 0.2106 \footnotesize \textcolor{gray}{ ± 0.0240 \textcolor{mygreen}{(↑5\%)}}  &  0.0125 \footnotesize \textcolor{gray}{ ± 0.0021 \textcolor{mygreen}{(↑8\%)}}  &  0.0104 \footnotesize \textcolor{gray}{ ± 0.0019 \textcolor{mygreen}{(↑7\%)}} \\
 & MFCC & 0.2609 \footnotesize \textcolor{gray}{ ± 0.0846 \textcolor{mygreen}{(↑15\%)}}  &  0.0182 \footnotesize \textcolor{gray}{ ± 0.0085 \textcolor{mygreen}{(↑24\%)}}  &  0.0141 \footnotesize \textcolor{gray}{ ± 0.0075 \textcolor{mygreen}{(↑18\%)}} \\
 & Music2Vec & 0.2993 \footnotesize \textcolor{gray}{ ± 0.0283 \textcolor{myred}{(↓3\%)}}  &  0.0250 \footnotesize \textcolor{gray}{ ± 0.0046 \textcolor{myred}{(↓3\%)}}  &  0.0207 \footnotesize \textcolor{gray}{ ± 0.0040 \textcolor{myred}{(↓2\%)}} \\
 & MERT & 0.3553 \footnotesize \textcolor{myred}{ ± 0.1317 \textcolor{mygreen}{(↑29\%)}}  &  0.0349 \footnotesize \textcolor{myred}{ ± 0.0262 \textcolor{mygreen}{(↑59\%)}}  &  0.0280 \footnotesize \textcolor{gray}{ ± 0.0206 \textcolor{mygreen}{(↑58\%)}} \\
 & MULE & 0.3947 \footnotesize \textcolor{myred}{ ± 0.1777 \textcolor{mygreen}{(↑70\%)}}  &  0.0442 \footnotesize \textcolor{myred}{ ± 0.0346 \textcolor{mygreen}{(↑191\%)}}  &  0.0350 \footnotesize \textcolor{myred}{ ± 0.0274 \textcolor{mygreen}{(↑176\%)}} \\
 & MuQ-MuLan & 0.4247 \footnotesize \textcolor{gray}{ ± 0.0186 \textcolor{mygreen}{(↑5\%)}}  &  0.0494 \footnotesize \textcolor{gray}{ ± 0.0039 \textcolor{mygreen}{(↑5\%)}}  &  0.0400 \footnotesize \textcolor{gray}{ ± 0.0030 \textcolor{mygreen}{(↑2\%)}} \\
 & MuQ & \underline{0.4410} \footnotesize \textcolor{gray}{ ± 0.0300 \textcolor{mygreen}{(↑61\%)}}  &  0.0535 \footnotesize \textcolor{gray}{ ± 0.0047 \textcolor{mygreen}{(↑149\%)}}  &  0.0443 \footnotesize \textcolor{gray}{ ± 0.0039 \textcolor{mygreen}{(↑150\%)}} \\
 & MusicFM & \underline{0.4427} \footnotesize \textcolor{gray}{ ± 0.0187 \textcolor{mygreen}{(↑81\%)}}  &  0.0530 \footnotesize \textcolor{gray}{ ± 0.0029 \textcolor{mygreen}{(↑205\%)}}  &  0.0439 \footnotesize \textcolor{gray}{ ± 0.0030 \textcolor{mygreen}{(↑207\%)}} \\
 & EncodecMAE & \underline{0.4449} \footnotesize \textcolor{gray}{ ± 0.0200 \textcolor{mygreen}{(↑4\%)}}  &  0.0536 \footnotesize \textcolor{gray}{ ± 0.0038 \textcolor{mygreen}{(↑8\%)}}  &  0.0450 \footnotesize \textcolor{gray}{ ± 0.0022 \textcolor{mygreen}{(↑7\%)}} \\
 & MusiCNN & \textbf{0.4722} \footnotesize \textcolor{gray}{ ± 0.0133 \textcolor{mygreen}{(↑11\%)}}  &  \textbf{0.0624} \footnotesize \textcolor{gray}{ ± 0.0021 \textcolor{mygreen}{(↑20\%)}}  &  \textbf{0.0520} \footnotesize \textcolor{gray}{ ± 0.0022 \textcolor{mygreen}{(↑19\%)}} \\

    \bottomrule
    \end{tabular}
    \label{tab:performance}
\end{table}

The structure of Table \ref{tab:performance} is the following: for each metric value, there are three numbers. Metric value ± margin of error (\% difference with a shuffled initialization).  

Each model was trained five times to estimate the mean and calculate the 95\% confidence interval plus three runs to estimate the mean of the shuffled initializations. Five runs were chosen to strike a balance between providing a reliable estimate of the true mean of the metric value for the model and the amount of computations needed for all comparisons. After all, the amount of computations grows fast with the number of reruns. In our experiment, for each of the ten PARs, we did five runs for four models for a normal initialization and then three runs for a shuffled initialization; we got a total of $10  \times (5 + 3) \times 4 =320$ experiment runs. Increasing the number of runs per PAR would result in even more experiment runs, making this research less feasible. We interpret the wide confidence intervals of some combinations of PARs and recommenders as an indicator of an instability of the particular combination which is less suited for practical use.

We use shuffled initialization to estimate the usefulness of a PAR for a recommender model. When we initialize the recommender model with PARs, we shuffle their order so that embeddings do not correspond to the item they were calculated for. This gives us the third number in a cell corresponding to a percentage increase (or decrease) of proper initialization compared to a shuffled one. Lower performance than shuffled initialization is unlikely, however, it can happen due to randomness and limited number of runs we used. Suppose the increase in performance is 0\% or a negative value. In that case, the model could not derive useful information from PARs and ignored them during learning. If the increase in metric is positive, PARs contain useful information that the model has a hard time ignoring with an incorrect order, which leads to a decreased quality in shuffled variant. This approach replaces the random initialization we used in the original publication of this paper \cite{Tamm_2024} and alleviates the problem of comparing a PAR-initialized model of one dimension size with a random-initialized model of the other dimension size. With shuffled initialization, the dimension size and statistical distribution of values across different dimensions remain the same for every PAR-Recommendation model pair.

The first observation is that, on average, more complicated recommendation models tend to give better performance. Across different pretrained representations, Shallow Net performs better than KNN, Bimodal better than Shallow Net, Hybrid better than Bimodal, and \mbox{BERT4Rec} performs similar to Hybrid, but allows to achieve the best result overall. This is expected but highlights the importance of choosing a model architecture with appropriate complexity.

If we look at the two worst-performing PARs across different models, we can find MFCC and Jukebox. 

It is not surprising to have MFCC at the bottom, because this is a handcrafted feature describing only the timbral aspect of sound. Seeing Jukebox at the bottom, on the other hand, is surprising because this PAR contains enough information for the task of music generation, but it doesn't seem to be so helpful for recommendation. The reason might be a large 4800 dimension size of Jukebox embeddings, which are the biggest in our experiment. We can indirectly support this hypothesis by studying the width of confidence intervals. This is especially true for the HitRate@50 of the Hybrid model, where the size of the confidence interval for Jukebox is very wide. We can confidently say that the model could not properly extract the information needed for the task. 

% Seeing MusicFM at low positions is surprising because it contradicts self-reported superiority in MIR tasks. We will discuss this comparison later in section \ref{mir}. We can also note that MusicFM is one of the few models that perform worse when the embeddings are in proper order (unshuffled). The reason for this behaviour is unclear and might suggest some technical problems with embeddings or the model weights we used.

For the middle four models performance varies. In the order of decreasing number of appearances in the middle-performance group: MERT~(4), MusicFM (3), Music2Vec~(3), MULE~(2), EncodecMAE~(2), MusiCNN~(1), MuQ-MuLan (1).

MERT seems to be performing well. However, the confidence intervals are rather wide: ±~0.026 with Shallow Net, ±~0.048 with Bimodal, and ±~0.13 with BERT4Rec. This indicates that its performance can probably be further improved with more elaborate models and more tuning. 

MusicFM shows medium performance for three recommenders out of four, reaching higher positions only with BERT4Rec. 

% These results are higher than previously reported in \cite{Tamm_2024}, where it was suggested that poor performance stems for some technical problems with embeddings, which turned out to be a problem with extraction code suggested in official repository.

Music2Vec notably does not give any advantage compared to shuffled initialization with BERT4Rec. This is similar with other PARs which have rather small gains. It seems that BERT4Rec is expressive enough to learn hidden representations mostly by itself, using provided PARs just as different initializations. Notably, the confidence interval for Music2Vec with a Hybrid recommender is extremely wide, indicating the instability of this combination. While this particular value should be interpreted with caution, it aligns with the performance of other recommenders with Music2Vec.

MULE seems to be performing normally, ending in the middle of the leaderboard across all models and scoring second place with Shallow Net. Similarly, the same applies to EncodecMAE.

MusiCNN falls short only with the Bimodal net, which is probably related to the particularites of the model that we will discuss later. 

Looking at the top-4 performing models, we can find MuQ (4), MuQ-MuLan (3), MusiCNN (3), MULE (2), EncodecMAE (2), Music2Vec (1), MusicFM (1).

Both MuQ models show top performance for hot recommendations with confidence intervals intersecting in most cases. Notably, MuQ-MuLan, which is enriched with text data, is not always higher than the base MuQ, and they perform similarly. MuQ models also have seen Music4All dataset during training, which could give them some advantage, although the models have not been trained for the task of recommendation, they just saw the same music when learning how to encode. 

MusiCNN is another well-performing model that shows the best overall performance, with 0.4722 HitRate with BERT4Rec. This is somewhat surprising because it is a relatively old model. Also, it is the only supervised model in our experiment, which might suggest similarities between tag prediction task and recommendations, and confirms findings from \cite{mccallum2022supervisedunsupervisedlearningaudio} about advantages of supervised learning. However, it is important to note that although other models have lower average metric values than MusiCNN, their confidence intervals overlap with MusiCNN's. They will probably diverge with more test runs, but their performance is generally comparable.

~\newline
\subsubsection{Recommender performance}~\newline

Another important question to ask is whether some recommenders are generally better suited for content information infusion. If a recommender allows different PARs to achieve their best performance, then it hints us a direction for more optimal hybrid architectures.

In order to asses this, we rearange Table \ref{tab:performance} into Table \ref{tab:par-by-hot-recs} by aggregating PAR HitRate@50 performance across recommenders. Every column has the best result highlighted. We can see that no PARs achieve their best performance with the Bimodal model and only two models work best with Shallow Net. However, there is no clear winner between a Hybrid model and BERT4Rec. Notably, both of these models are more complex than Shallow Net or Bimodal. This indicates that for the hot-start scenario more elaborate models might perform better than simpler models.

\begin{table}[htbp]
    \centering
    \caption{HitRate@50 Performance of PARs with different Recommenders for hot-start scenario}
    \resizebox{\textwidth}{!}{%
    \begin{tabular}{lccccccccccc}
    \toprule
Recommender & MFM & MFCC & M2V & JBX & MCNN & MERT & EMAE & MULE & MuQ & MuLan \\
\midrule
Bm Average   & 0.1698 & 0.0435 & 0.2850 & 0.1157 & 0.1657 & 0.2462 & 0.2551 & 0.2271 & 0.3794 & 0.3975 \\
Shallow Net  & 0.3217 & 0.2483 & \textbf{0.3200} & \textbf{0.2982} & 0.3581 & 0.3264 & 0.3319 & 0.3735 & 0.3895 & 0.3730 \\
Hybrid       & 0.3067 & \textbf{0.3021} & 0.3025 & 0.0775 & 0.4323 & \textbf{0.3680} & 0.3804 & \textbf{0.4071} & 0.4239 & \textbf{0.4450} \\
BERT4Rec     & \textbf{0.4427} & 0.2609 & 0.2993 & 0.2106 & \textbf{0.4722} & 0.3553 & \textbf{0.4449} & 0.3947 & \textbf{0.4410} & 0.4247 \\
\bottomrule
    \end{tabular}%
    }
    \label{tab:par-by-hot-recs}
\end{table}

\subsubsection{Bimodal performance}~\newline
% \vspace{-2.7em}
\begin{table}[htbp]
    \centering
    \caption{Performance of different modalities trained with a Bimodal contrastive approach. Collaborative column refers to a projection of embeddings from the ELSA model, Content is a projection of the respective PAR, and the Average projection is the mean of the two. Bold indicates the highest mean. Underlined values indicate methods whose 95\% confidence interval overlaps with that of the best-performing method.}
    \begin{tabular}{llllccc}
    \toprule
    \multicolumn{1}{c}{Embeddings} & \multicolumn{3}{c}{Bimodal HitRate@50} \\
    \midrule
    \multicolumn{1}{l}{} & \multicolumn{1}{c}{Collaborative} & \multicolumn{1}{c}{Average} & \multicolumn{1}{c}{Content} \\
    \midrule
MFCC & 0.2565 \footnotesize \textcolor{gray}{ ± 0.0222 \textcolor{mygreen}{(↑243\%)}} & 0.0435 \footnotesize \textcolor{gray}{ ± 0.0054 \textcolor{mygreen}{(↑80\%)}} & 0.0434 \footnotesize \textcolor{gray}{ ± 0.0052 \textcolor{mygreen}{(↑79\%)}}  \\
 Jukebox & 0.3861 \footnotesize \textcolor{gray}{ ± 0.0078 \textcolor{mygreen}{(↑416\%)}} & 0.1157 \footnotesize \textcolor{gray}{ ± 0.0464 \textcolor{mygreen}{(↑361\%)}}  & 0.1106 \footnotesize \textcolor{gray}{ ± 0.0463 \textcolor{mygreen}{(↑346\%)}}  \\
 MusiCNN & 0.3273 \footnotesize \textcolor{gray}{ ± 0.0086 \textcolor{mygreen}{(↑257\%)}} & 0.1657 \footnotesize \textcolor{gray}{ ± 0.0086 \textcolor{mygreen}{(↑511\%)}}  & 0.1442 \footnotesize \textcolor{gray}{ ± 0.0132 \textcolor{mygreen}{(↑489\%)}}  \\
 MusicFM & \underline{0.3967} \footnotesize \textcolor{gray}{  ± 0.0114\textcolor{mygreen}{(↑396\%)}}  & 0.1698 \footnotesize \textcolor{gray}{  ± 0.0353 \textcolor{mygreen}{(↑541\%) }} & 0.1178 \footnotesize \textcolor{gray}{ ± 0.0332  \textcolor{mygreen}{(↑350\%)}}  \\
 MULE & 0.3888 \footnotesize \textcolor{gray}{ ± 0.0123 \textcolor{mygreen}{(↑222\%)}}  & 0.2271 \footnotesize \textcolor{gray}{ ± 0.0248 \textcolor{mygreen}{(↑691\%)}} & \underline{0.2147} \footnotesize \textcolor{gray}{ ± 0.0276 \textcolor{mygreen}{(↑678\%)}}  \\
 MERT & 0.3663 \footnotesize \textcolor{gray}{ ± 0.0168 \textcolor{mygreen}{(↑350\%)}}  & 0.2462 \footnotesize \textcolor{gray}{ ± 0.0481 \textcolor{mygreen}{(↑851\%)}}  & 0.1095 \footnotesize \textcolor{gray}{ ± 0.0379 \textcolor{mygreen}{(↑324\%)}}   \\
 EncodecMAE & 0.3586 \footnotesize \textcolor{gray}{ ± 0.0189 \textcolor{mygreen}{(↑329\%)}} & 0.2551 \footnotesize \textcolor{gray}{ ± 0.0111 \textcolor{mygreen}{(↑855\%)}} & 0.1265 \footnotesize \textcolor{gray}{ ± 0.0151 \textcolor{mygreen}{(↑443\%)}}  \\
 Music2Vec & 0.2952 \footnotesize \textcolor{gray}{ ± 0.0143 \textcolor{mygreen}{(↑287\%)}}   & 0.2850 \footnotesize \textcolor{gray}{ ± 0.0168 \textcolor{mygreen}{(↑652\%)}} & 0.1016 \footnotesize \textcolor{gray}{ ± 0.0112 \textcolor{mygreen}{(↑294\%)}}  \\
 MuQ & \textbf{0.4145} \footnotesize \textcolor{gray}{ ± 0.0121 \textcolor{mygreen}{(↑272\%)}}  & \underline{0.3794} \footnotesize \textcolor{gray}{ ± 0.0115 \textcolor{mygreen}{(↑1418\%)}} & \textbf{0.2163} \footnotesize \textcolor{gray}{ ± 0.0171 \textcolor{mygreen}{(↑857\%)}}  \\
 MuQ-MuLan & \underline{0.3995} \footnotesize \textcolor{gray}{ ± 0.0192 \textcolor{mygreen}{(↑467\%)}}  & \textbf{0.3975} \footnotesize \textcolor{gray}{ ± 0.0173 \textcolor{mygreen}{(↑1029\%)}} & 0.1720 \footnotesize \textcolor{gray}{ ± 0.0109 \textcolor{mygreen}{(↑588\%)}}  \\
    \bottomrule
    \end{tabular}
    \label{tab:bimodal}
\end{table}

Now let us look at Table~\ref{tab:bimodal} and discuss the options we have for the Bimodal net. When we train a Bimodal net, we take both collaborative and content vectors as input, project them to a new space and require them to be close to each other. This naturally allows us to use one of the three projections: collaborative, content or their mean. We took the average version as the main one for the hot test because it is supposed to capture both modalities well. However, collaborative projection works better than content and average ones. The reason is because ELSA embeddings alone allow us to get a superior 0.558 HitRate without any content information added. That is, when we train a Bimodal net, we cannot achieve the same quality as the original collaborative version, even with the collaborative projection. 

While this might seem counterintuitive, adding extra modalities does not necessarily lead to an improved quality. Recent research \cite{zhou2025does} shows that a purely collaborative UserKNN model can outperform some multi-modal models. Moreover, using fewer modalities or only interaction data can sometimes work better than multi-modal versions of the same models. Another example of this can be found in \cite{spillo2025seef}
where the direct expansion of BPR into VBPR using audio information performs worse than the collaborative variant on the Last.fm-2k dataset.

Content information helps solve the cold-start problem, but hybrid models can be outperformed by their collaborative counterparts or other CF models in the hot scenario. As such, Bimodal embeddings may have some other good properties, but they do not allow achieving the best HitRate values in our experiment.

~\newline
\subsubsection{General performance}~\newline

\begin{table}[htbp]
    \centering
    \caption{Best-performing version of every hybrid recommendation model and their collaborative versions for the hot test}
    \begin{tabular}{lcccccc}
    \toprule
    Model & Type & HitRate@50 & Recall@50 & NDCG@50 \\
    \midrule
    KNN & Content & 0.0370  &  0.0012 &  0.0010 \\
    Shallow Net & Collaborative & 0.0506   &  0.0019 &  0.0017 \\
    Bimodal Content & Hybrid & 0.2163 & 0.0173 &  0.0120 \\
    PopRec & Collaborative &  0.2292 & 0.0149 & 0.0129 \\
    Shallow Net & Hybrid &  0.3895   &  0.0412 &  0.0308 \\
    Bimodal Average & Hybrid &  0.3975   &  0.0358 &  0.0276 \\
    Bimodal Collaborative & Hybrid &  0.4145 &  0.0383 &  0.0298 \\
    BERT4Rec & Collaborative & 0.4349   &  0.0559   & 0.0470\\ 
    Hybrid & Hybrid &  0.4450   & 0.0485 &  0.0390 \\
    BERT4Rec & Hybrid &   0.4722  &  0.0624 &  0.0520 \\
    ELSA & Collaborative &   0.5580   &  0.0823 &  0.0631  \\

    \bottomrule
    \end{tabular}
    \label{tab:elsa}
\end{table}

Continuing with Table~\ref{tab:elsa}, let us discuss the performance of the models we tested using the best PAR for each model. We used ELSA collaborative embeddings to train the Bimodal representation, so we added its separate performance into the comparison. We also added a popularity-based recommender to give a reference point to the performance of other variants. The type column signifies whether the model used content, collaborative information, or both for training. Collaborative versions of Shallow Net and BERT4Rec use the same architecture we described earlier, but instead of initializing them with PARs, we initialize their weights randomly, thus using only collaborative information for their training.

We can see that in the hot scenario, our recommendation models cannot beat the pure collaborative model ELSA. It is important to stress the difference between improvements of content initializations over the same recommendation model and a different model. We can see that content initializations improve collaborative versions of Shallow Net and BERT4Rec. However, this does not directly translate to the best overall performance, highlighting the importance of using strong baselines. In hot-start scenario, this way of adding content information might not lead to the direct improvement of accuracy metrics over strong purely collaborative models like ELSA. However, it has the advantage of a unified model for both hot and cold scenarios. Moreover, we do not study beyond-accuracy metrics in this work, which may differ for hybrid and collaborative variants. In this paper, we focus on evaluating different PARs, and so we leave the deeper comparison of collaborative and hybrid models, as well as the search for the optimal hybrid model architecture, for future work.

\newpage
\subsection{Cold Test}
Following the structure of the previous section, we start by looking at the performance of the KNN model in Table \ref{tab:knn-cold}.

The first thing we should outline is the difference in scale. In the cold scenario, the values are much larger than those we have seen in the hot one. They should not be directly compared to each other because, for the cold start, we only use a much smaller set of 589 cold items as candidates for recommendations, which naturally makes all the metrics higher.

However, when we study the order of the models, we can see  MERT was in the lower part of the rankings, and here it rose to top-1, whereas MusicFM was top-1 in the hot-start and got a lower position here. Moreover, other models also changed places; MuQ-MuLan, MuQ, and EncodecMAE got lower positions.

\begin{table}[htbp]
    \centering
    \caption{Performance of pretrained embeddings in the cold-start scenario using KNN}
    \begin{tabular}{lcccccc}
    \toprule
    Embeddings & HitRate@20 & Recall@20 & NDCG@20 \\
    \midrule
    MuQ  &     0.1726    &   0.0487  &   0.0191 \\
    Jukebox  &     0.1832    &   0.0500  &   0.0180 \\
    MULE   &       0.2242  &   0.0666 &  0.0258 \\
    MFCC  &        0.2423    &   0.0692  &   0.0247 \\
    MusicFM  &  0.2453    &   0.0697  &   0.0316 \\
    Music2Vec  &   0.2482    &   0.0650  &   0.0302 \\
    MuQ-MuLan  &    0.2502   &  0.0686  &   0.0275 \\
    EncodecMAE  &  0.2504    &   0.0671  &   0.0254 \\
    MusiCNN  &     0.2662    &   \textbf{0.1189}  &   \textbf{0.0496} \\
    MERT  &      \textbf{0.2829}    &   0.0840  &   0.0320 \\
    \bottomrule
    \end{tabular}
    \label{tab:knn-cold}
\end{table}

\begin{table}[htbp]
    \centering
    \caption{Comparison of performance of PARs using different recommenders in the cold-start scenario. Cell format: metric value ± margin of error (\% difference with a shuffled initialization). Bold indicates the highest mean. Underlined values indicate methods whose 95\% CI overlaps with that of the best-performing method. Red CI indicate high variance across runs. For BERT4Rec, all CI overlap due to high variance.}
    \begin{tabular}{cllllccc}
    \toprule
    & \multicolumn{1}{l}{Embeddings} & \multicolumn{1}{c}{HitRate@20} & \multicolumn{1}{c}{Recall@20} & \multicolumn{1}{c}{NDCG@20} \\

    \midrule
    \multirow{10}{*}{\rotatebox{90}{Shallow Net}}
 & Music2Vec & 0.4360 \footnotesize \textcolor{gray}{ ± 0.0312 \textcolor{mygreen}{(↑49\%)}}  &  0.1502 \footnotesize \textcolor{gray}{ ± 0.0177 \textcolor{mygreen}{(↑88\%)}}  &  0.0642 \footnotesize \textcolor{gray}{ ± 0.0076 \textcolor{mygreen}{(↑106\%)}} \\
 & MERT & 0.4651 \footnotesize \textcolor{gray}{ ± 0.0637 \textcolor{mygreen}{(↑49\%)}}  &  0.1747 \footnotesize \textcolor{gray}{ ± 0.0354 \textcolor{mygreen}{(↑105\%)}}  &  0.0799 \footnotesize \textcolor{gray}{ ± 0.0181 \textcolor{mygreen}{(↑128\%)}} \\
 & MFCC & 0.4866 \footnotesize \textcolor{gray}{ ± 0.0261 \textcolor{mygreen}{(↑57\%)}}  &  0.1923 \footnotesize \textcolor{gray}{ ± 0.0262 \textcolor{mygreen}{(↑123\%)}}  &  0.0910 \footnotesize \textcolor{gray}{ ± 0.0128 \textcolor{mygreen}{(↑179\%)}} \\
  & MusicFM & \underline{0.4992} \footnotesize \textcolor{gray}{ ± 0.0499 \textcolor{mygreen}{ (↑68\%)}}  &  \underline{0.1986} \footnotesize \textcolor{gray}{ ± 0.0324  \textcolor{mygreen}{(↑141\%)}}  &  \underline{0.0922} \footnotesize \textcolor{gray}{± 0.0197 \textcolor{mygreen}{(↑178\%)}} \\
 & EncodecMAE & 0.4993 \footnotesize \textcolor{gray}{ ± 0.0432 \textcolor{mygreen}{(↑79\%)}}  &  \underline{0.2012} \footnotesize \textcolor{gray}{ ± 0.0289 \textcolor{mygreen}{(↑163\%)}}  &  \underline{0.0921} \footnotesize \textcolor{gray}{ ± 0.0221 \textcolor{mygreen}{(↑214\%)}} \\
 & MusiCNN & 0.5038 \footnotesize \textcolor{gray}{ ± 0.0249 \textcolor{mygreen}{(↑51\%)}}  &  \underline{0.2089} \footnotesize \textcolor{gray}{ ± 0.0177 \textcolor{mygreen}{(↑125\%)}}  &  0.0944 \footnotesize \textcolor{gray}{ ± 0.0086 \textcolor{mygreen}{(↑158\%)}} \\
 & Jukebox & \underline{0.5071} \footnotesize \textcolor{gray}{ ± 0.0547 \textcolor{mygreen}{(↑56\%)}}  &  \underline{0.2059} \footnotesize \textcolor{gray}{ ± 0.0413 \textcolor{mygreen}{(↑129\%)}}  &  \underline{0.0953} \footnotesize \textcolor{gray}{ ± 0.0268 \textcolor{mygreen}{(↑160\%)}} \\
 & MuQ-MuLan & 0.5172 \footnotesize \textcolor{gray}{ ± 0.0232 \textcolor{mygreen}{(↑72\%)}}  &  \underline{0.2235} \footnotesize \textcolor{gray}{ ± 0.0088 \textcolor{mygreen}{(↑177\%)}}  &  0.1025 \footnotesize \textcolor{gray}{ ± 0.0026 \textcolor{mygreen}{(↑219\%)}} \\
 & MULE & \underline{0.5180} \footnotesize \textcolor{gray}{ ± 0.0385 \textcolor{mygreen}{(↑69\%)}}  &  \underline{0.2113} \footnotesize \textcolor{gray}{ ± 0.0243 \textcolor{mygreen}{(↑157\%)}}  &  \underline{0.0974} \footnotesize \textcolor{gray}{ ± 0.0121 \textcolor{mygreen}{(↑192\%)}} \\
 & MuQ & \textbf{0.5597} \footnotesize \textcolor{gray}{ ± 0.0166 \textcolor{mygreen}{(↑75\%)}}  &  \textbf{0.2435} \footnotesize \textcolor{gray}{ ± 0.0180 \textcolor{mygreen}{(↑166\%)}}  &  \textbf{0.1130} \footnotesize \textcolor{gray}{ ± 0.0074 \textcolor{mygreen}{(↑190\%)}} \\
    
    \midrule
    \addlinespace[0.7em]
    \midrule

    \multirow{10}{*}{\rotatebox{90}{Bimodal Content}} 
 & MFCC & 0.4742 \footnotesize \textcolor{gray}{ ± 0.0162 \textcolor{mygreen}{(↑67\%)}}  &  0.2148 \footnotesize \textcolor{gray}{ ± 0.0139 \textcolor{mygreen}{(↑208\%)}}  &  0.1016 \footnotesize \textcolor{gray}{ ± 0.0140 \textcolor{mygreen}{(↑281\%)}} \\
 & MusiCNN & 0.5686 \footnotesize \textcolor{gray}{ ± 0.0323 \textcolor{mygreen}{(↑88\%)}}  &  0.2862 \footnotesize \textcolor{gray}{ ± 0.0293 \textcolor{mygreen}{(↑252\%)}}  &  0.1279 \footnotesize \textcolor{gray}{ ± 0.0181 \textcolor{mygreen}{(↑260\%)}} \\
 & MULE & \underline{0.6022} \footnotesize \textcolor{gray}{ ± 0.0731 \textcolor{mygreen}{(↑117\%)}}  &  0.3160 \footnotesize \textcolor{gray}{ ± 0.0648 \textcolor{mygreen}{(↑305\%)}}  &  \underline{0.1512} \footnotesize \textcolor{gray}{ ± 0.0415 \textcolor{mygreen}{(↑406\%)}} \\
 & Music2Vec & 0.6106 \footnotesize \textcolor{gray}{ ± 0.0069 \textcolor{mygreen}{(↑110\%)}}  &  0.3087 \footnotesize \textcolor{gray}{ ± 0.0152 \textcolor{mygreen}{(↑275\%)}}  &  0.1378 \footnotesize \textcolor{gray}{ ± 0.0112 \textcolor{mygreen}{(↑342\%)}} \\
 & EncodecMAE & 0.6202 \footnotesize \textcolor{gray}{ ± 0.0251 \textcolor{mygreen}{(↑78\%)}}  &  0.3410 \footnotesize \textcolor{gray}{ ± 0.0271 \textcolor{mygreen}{(↑239\%)}}  &  0.1591 \footnotesize \textcolor{gray}{ ± 0.0171 \textcolor{mygreen}{(↑296\%)}} \\
 & MERT & \underline{0.6405} \footnotesize \textcolor{gray}{ ± 0.0500 \textcolor{mygreen}{(↑130\%)}}  &  \underline{0.3590} \footnotesize \textcolor{gray}{ ± 0.0547 \textcolor{mygreen}{(↑352\%)}}  &  \underline{0.1716} \footnotesize \textcolor{gray}{ ± 0.0350 \textcolor{mygreen}{(↑426\%)}} \\
 & MuQ-MuLan & 0.6468 \footnotesize \textcolor{gray}{ ± 0.0201 \textcolor{mygreen}{(↑133\%)}}  &  0.3577 \footnotesize \textcolor{gray}{ ± 0.0142 \textcolor{mygreen}{(↑378\%)}}  &  0.1632 \footnotesize \textcolor{gray}{ ± 0.0101 \textcolor{mygreen}{(↑446\%)}} \\
 & MuQ & \underline{0.6517} \footnotesize \textcolor{gray}{ ± 0.0263 \textcolor{mygreen}{(↑96\%)}}  &  \underline{0.3740} \footnotesize \textcolor{gray}{ ± 0.0227 \textcolor{mygreen}{(↑301\%)}}  &  \underline{0.1805} \footnotesize \textcolor{gray}{ ± 0.0158 \textcolor{mygreen}{(↑322\%)}} \\
& MusicFM & \underline{0.6599} \footnotesize \textcolor{gray}{  ± 0.0329  \textcolor{mygreen}{(↑117\%)}}  &  \underline{0.3760} \footnotesize \textcolor{gray}{ ± 0.0421 \textcolor{mygreen}{(↑323\%)}}  & \underline{0.1737} \footnotesize \textcolor{gray}{ ± 0.0205 \textcolor{mygreen}{(↑347\%)}} \\
 & Jukebox & \textbf{0.6788} \footnotesize \textcolor{gray}{ ± 0.0104 \textcolor{mygreen}{(↑122\%)}}  &  \textbf{0.3966} \footnotesize \textcolor{gray}{ ± 0.0093 \textcolor{mygreen}{(↑394\%)}}  &  \textbf{0.1862} \footnotesize \textcolor{gray}{ ± 0.0082 \textcolor{mygreen}{(↑457\%)}} \\

    \midrule
    \addlinespace[0.7em]
    \midrule
    \multirow{10}{*}{\rotatebox{90}{Hybrid}}
 & Jukebox & \underline{0.4469} \footnotesize \textcolor{myred}{ ± 0.2113} \footnotesize \textcolor{gray}{ \textcolor{mygreen}{(↑38\%)}}  &  \underline{0.2113} \footnotesize \textcolor{myred}{ ± 0.1844} \footnotesize \textcolor{gray}{ \textcolor{mygreen}{(↑161\%)}}  &  0.0858 \footnotesize \textcolor{myred}{ ± 0.0763} \footnotesize \textcolor{gray}{ \textcolor{mygreen}{(↑171\%)}} \\
 & MULE & 0.5099 \footnotesize \textcolor{gray}{ ± 0.0342 \textcolor{mygreen}{(↑72\%)}}  &  0.2148 \footnotesize \textcolor{gray}{ ± 0.0300 \textcolor{mygreen}{(↑161\%)}}  &  0.1086 \footnotesize \textcolor{gray}{ ± 0.0182 \textcolor{mygreen}{(↑229\%)}} \\
 & MusiCNN & 0.5270 \footnotesize \textcolor{gray}{ ± 0.0266 \textcolor{mygreen}{(↑75\%)}}  &  0.2263 \footnotesize \textcolor{gray}{ ± 0.0166 \textcolor{mygreen}{(↑177\%)}}  &  0.1080 \footnotesize \textcolor{gray}{ ± 0.0103 \textcolor{mygreen}{(↑223\%)}} \\
 & MuQ & 0.5396 \footnotesize \textcolor{gray}{ ± 0.0320 \textcolor{mygreen}{(↑102\%)}}  &  0.2362 \footnotesize \textcolor{gray}{ ± 0.0265 \textcolor{mygreen}{(↑243\%)}}  &  0.1174 \footnotesize \textcolor{gray}{ ± 0.0161 \textcolor{mygreen}{(↑316\%)}} \\
 & Music2Vec & \underline{0.5547} \footnotesize \textcolor{myred}{ ± 0.2208} \footnotesize \textcolor{gray}{ \textcolor{mygreen}{(↑59\%)}}  &  \underline{0.2668} \footnotesize \textcolor{myred}{ ± 0.1729} \footnotesize \textcolor{gray}{ \textcolor{mygreen}{(↑202\%)}}  &  \underline{0.1236} \footnotesize \textcolor{myred}{ ± 0.0828} \footnotesize \textcolor{gray}{ \textcolor{mygreen}{(↑272\%)}} \\
 & MFCC & 0.5659 \footnotesize \textcolor{gray}{ ± 0.0588 \textcolor{mygreen}{(↑89\%)}}  &  0.2715 \footnotesize \textcolor{gray}{ ± 0.0452 \textcolor{mygreen}{(↑246\%)}}  &  0.1223 \footnotesize \textcolor{gray}{ ± 0.0142 \textcolor{mygreen}{(↑274\%)}} \\
 & EncodecMAE & 0.6019 \footnotesize \textcolor{gray}{ ± 0.0399 \textcolor{mygreen}{(↑102\%)}}  &  0.3130 \footnotesize \textcolor{gray}{ ± 0.0415 \textcolor{mygreen}{(↑288\%)}}  &  0.1514 \footnotesize \textcolor{gray}{ ± 0.0239 \textcolor{mygreen}{(↑362\%)}} \\
 & MERT & 0.6257 \footnotesize \textcolor{gray}{ ± 0.0089 \textcolor{mygreen}{(↑128\%)}}  &  0.3509 \footnotesize \textcolor{gray}{ ± 0.0106 \textcolor{mygreen}{(↑412\%)}}  &  0.1676 \footnotesize \textcolor{gray}{ ± 0.0079 \textcolor{mygreen}{(↑535\%)}} \\
 & MuQ-MuLan & \underline{0.6697} \footnotesize \textcolor{gray}{ ± 0.0385 \textcolor{mygreen}{(↑134\%)}}  &  \underline{0.3615} \footnotesize \textcolor{gray}{ ± 0.0437 \textcolor{mygreen}{(↑399\%)}}  &  \underline{0.1816} \footnotesize \textcolor{gray}{ ± 0.0212 \textcolor{mygreen}{(↑533\%)}} \\
 & MusicFM & \textbf{0.6727} \footnotesize \textcolor{gray}{± 0.0208  \textcolor{mygreen}{(↑169\%)}}  &  \textbf{0.3917} \footnotesize \textcolor{gray}{ ± 0.0161 \textcolor{mygreen}{(↑460\%)}}  &  \textbf{0.1871} \footnotesize \textcolor{gray}{ ± 0.0114 \textcolor{mygreen}{(↑418\%) }} \\

    \midrule
    \addlinespace[0.7em]
    \midrule
        \multirow{10}{*}{\rotatebox{90}{BERT4Rec}}
 & Jukebox & 0.3062 \footnotesize \textcolor{gray}{ ± 0.0719 \textcolor{myred}{(↓14\%)}}  &  0.1037 \footnotesize \textcolor{gray}{ ± 0.0608 \textcolor{myred}{(↓8\%)}}  &  0.0368 \footnotesize \textcolor{gray}{ ± 0.0180 \textcolor{myred}{(↓23\%)}} \\
 & Music2Vec & 0.3914 \footnotesize \textcolor{myred}{ ± 0.1195} \footnotesize \textcolor{gray}{ \textcolor{mygreen}{(↑49\%)}}  &  0.1279 \footnotesize \textcolor{myred}{ ± 0.0688} \footnotesize \textcolor{gray}{ \textcolor{mygreen}{(↑83\%)}}  &  0.0512 \footnotesize \textcolor{myred}{ ± 0.0307} \footnotesize \textcolor{gray}{ \textcolor{mygreen}{(↑91\%)}} \\
 & EncodecMAE & 0.4170 \footnotesize \textcolor{gray}{ ± 0.0415 \textcolor{mygreen}{(↑30\%)}}  &  0.1624 \footnotesize \textcolor{gray}{ ± 0.0280 \textcolor{mygreen}{(↑86\%)}}  &  0.0677 \footnotesize \textcolor{gray}{ ± 0.0172 \textcolor{mygreen}{(↑90\%)}} \\
 & MERT & 0.4170 \footnotesize \textcolor{gray}{ ± 0.0477 \textcolor{mygreen}{(↑27\%)}}  &  0.1468 \footnotesize \textcolor{gray}{ ± 0.0346 \textcolor{mygreen}{(↑74\%)}}  &  0.0678 \footnotesize \textcolor{gray}{ ± 0.0210 \textcolor{mygreen}{(↑99\%)}} \\
 & MFCC & 0.4367 \footnotesize \textcolor{gray}{ ± 0.0684 \textcolor{mygreen}{(↑30\%)}}  &  0.1780 \footnotesize \textcolor{gray}{ ± 0.0380 \textcolor{mygreen}{(↑106\%)}}  &  0.0750 \footnotesize \textcolor{gray}{ ± 0.0135 \textcolor{mygreen}{(↑130\%)}} \\
 & MusiCNN & 0.4589 \footnotesize \textcolor{gray}{ ± 0.0801 \textcolor{mygreen}{(↑38\%)}}  &  0.1888 \footnotesize \textcolor{gray}{ ± 0.0503 \textcolor{mygreen}{(↑107\%)}}  &  0.0780 \footnotesize \textcolor{gray}{ ± 0.0245 \textcolor{mygreen}{(↑104\%)}} \\
 & MuQ-MuLan & 0.4846 \footnotesize \textcolor{gray}{ ± 0.0691 \textcolor{mygreen}{(↑57\%)}}  &  0.2047 \footnotesize \textcolor{gray}{ ± 0.0400 \textcolor{mygreen}{(↑139\%)}}  &  0.0877 \footnotesize \textcolor{gray}{ ± 0.0187 \textcolor{mygreen}{(↑163\%)}} \\
  & MusicFM & 0.4898 \footnotesize \textcolor{gray}{ ± 0.0383 \textcolor{mygreen}{(↑74\%)}}  &  0.2111 \footnotesize \textcolor{gray}{ ± 0.0265  \textcolor{mygreen}{(↑160\%)}}  &  0.0824 \footnotesize \textcolor{gray}{ ± 0.0163  \textcolor{mygreen}{(↑163\%)}} \\
 & MuQ & 0.5117 \footnotesize \textcolor{gray}{ ± 0.0892 \textcolor{mygreen}{(↑56\%)}}  &  0.2243 \footnotesize \textcolor{gray}{ ± 0.0564 \textcolor{mygreen}{(↑139\%)}}  &  0.0910 \footnotesize \textcolor{gray}{ ± 0.0363 \textcolor{mygreen}{(↑152\%)}} \\
 & MULE & \textbf{0.5118} \footnotesize \textcolor{myred}{ ± 0.1788} \footnotesize \textcolor{gray}{ \textcolor{mygreen}{(↑49\%)}}  &  \textbf{0.2393} \footnotesize \textcolor{myred}{ ± 0.1745} \footnotesize \textcolor{gray}{ \textcolor{mygreen}{(↑163\%)}}  &  \textbf{0.1079} \footnotesize \textcolor{myred}{ ± 0.0751} \footnotesize \textcolor{gray}{ \textcolor{mygreen}{(↑219\%)}} \\
    \bottomrule
    \end{tabular}
    \label{tab:performance-cold}
\end{table}
~\newline 
\subsubsection{PAR performance}~\newline 

Now, we turn our attention to Table \ref{tab:performance-cold}. The first thing we notice is that the ranking of recommender models differs from that observed for hot-item recommendations when sorting by the performance of the top-ranking PARs for each model. In the hot scenario, if we rank models by performance increasingly, we would get Shallow Net, Bimodal, Hybrid and BERT4Rec; here, we have BERT4Rec, Shallow Net, Hybrid and Bimodal. 

Interestingly, even though the Bimodal net struggled in the hot scenario in our experiment, it performed best in the cold scenario. Note that we use the Bimodal Content variant for the cold scenario since a collaborative vector is not available for the cold items. It seems that infusing content projections with content information through contrastive loss is especially useful for the cold-start scenario.
% Moreover, if we look at the worst performing PAR for the Bimodal net, the MusicFM, it performs considerably better with this recommender model than with others.

The second thing to note is that in the cold scenario, results are less stable, with many PARs having overlapping confidence intervals with the best performing option across all the recommender models. Notably, for BERT4Rec, all confidence intervals overlap due to the high variance of MULE, and no statistically reliable ranking
can be established. This means the exact rankings are less clear-cut, but we can still study the general picture. 

The lowest-performing places are shared by most models at least one time: Music2Vec~(3), MERT~(2),  MFCC~(2), MusiCNN~(2), Jukebox~(2), MULE~(2), MusicFM~(1).

When we look at the middle three positions, we can see that different models end up in this area: EncodecMAE~(3), MusiCNN~(2), MFCC~(2), MULE~(1), Music2Vec~(1), MuQ-Mulan~(2), MERT~(1), Jukebox~(1). Notably, MusiCNN, one of the top performers in the hot scenario, showed worse performance in the cold scenario and did not reach top-3.

Top-3 places are shared by MusicFM~(3), MuQ~(3), MuQ-MuLan~(2), MULE~(2), Jukebox~(1), MERT~(1). Similarly to the hot scenario, MuQ and MuQ-MuLan consistently perform well. Interestingly, Jukebox, which was in a lower position in the hot scenario, now rose higher, allowing it to achieve the best overall result. However, the increase over shuffled initialization is negative for BERT4Rec, which indicates that the model could not put content information to good use. Notably, MusicFM performs much better in the cold scenario than it did in the hot scenario and achieves comparative performance to a much larger Jukebox.

We should also note that MuQ-MuLan performs best with the Hybrid recommender across both hot and cold scenarios, making this PAR a strong candidate for a content backend within a unified model. 

While some PARs, like MuQ-MuLan, show stable performance across both hot and cold scenarios, others may be more useful for a particular scenario. As such, MusiCNN performs best in the hot scenario, while Jukebox and MusicFM are most useful in the cold scenario.

~\newline 
\subsubsection{Recommender performance}~\newline 

When we rearange the results for the best performing recommenders into Table \ref{tab:par-by-cold-recs} we can see that in cold-start scenario Bimodal recommender almost universally achieves best results for different PARs. This is in a stark contrast with the hot-start scenario in Table \ref{tab:par-by-hot-recs} where better performing recommenders were Hybrid and BERT4Rec. This suggests that it maybe beneficial to employ different recommenders for hot and cold scenarios.

\begin{table}[htbp]
    \centering
    \caption{HitRate@20 Performance of PARs with different Recommenders for cold-start scenario}
    \resizebox{\textwidth}{!}{%
    \begin{tabular}{lcccccccccc}
    \toprule
Recommender & MFM & MFCC & M2V & JBX & MCNN & MERT & EMAE & MULE & MuQ & MuLan \\
\midrule
Bm Content   & 0.6599 & 0.4742 & \textbf{0.6106} & \textbf{0.6788} & \textbf{0.5686} & \textbf{0.6405} & \textbf{0.6202} & \textbf{0.6022} & \textbf{0.6517} & 0.6468 \\
Shallow Net  & 0.4992 & 0.4866 & 0.4360 & 0.5071 & 0.5038 & 0.4651 & 0.4993 & 0.5180 & 0.5597 & 0.5172 \\
Hybrid       & \textbf{0.6727} & \textbf{0.5659} & 0.5547 & 0.4469 & 0.5270 & 0.6257 & 0.6019 & 0.5099 & 0.5396 & \textbf{0.6697} \\
BERT4Rec     & 0.4898 & 0.4367 & 0.3914 & 0.3062 & 0.4589 & 0.4170 & 0.4170 & 0.5118 & 0.5117 & 0.4846 \\
\bottomrule
    \end{tabular}%
    }
    \label{tab:par-by-cold-recs}
\end{table}

\subsection{Comparison to MIR results}
\label{mir}

\begin{table}
    \centering
\caption{Comparison of backend models applied to different MIR and MRS tasks. Results for MIR are taken from the respective papers of each model; results for Hot and Cold recommendations are the best HitRate values for corresponding models from Table \ref{tab:performance} and Table \ref{tab:performance-cold}, respectively.}
\label{tab:mir}
    \begin{tabular}{cccccc} 
         Model&  Tags&  Genres&  Key& Hot Recs & Cold Recs\\ 
         MusicFM&  \textbf{0.924}&  —&  \textbf{0.674}& 0.443  & 0.673\\ 
         Music2Vec&  0.895&  0.766&  0.501& 0.320 & 0.611\\ 
         MERT&  0.913&  0.793&  0.656& 0.368 & 0.641\\ 
         EncodecMAE&  —&  \textbf{0.878}&  —& 0.445 & 0.620\\ 
         Jukebox&  0.915&  0.797&  0.667& 0.298 & \textbf{0.679}\\ 
         MusiCNN& 0.906& 0.790& 0.128&\textbf{0.472} & 0.569\\
         MuQ& 0.914& 0.856& 0.650& 0.441 & 0.652\\
         MULE& 0.917& 0.835& 0.286 & 0.407 & 0.602\\
         MFCC& 0.858 & 0.448 & 0.146 & 0.302 & 0.566\\
    \end{tabular}
\end{table}
In this section, we aggregate the results so far and compare them with self-reported performance on the MIR tasks for each model, to examine relative ordering and determine whether top performers in one task also happen to be top performers in the other.

We used self-reported data from corresponding papers to compare our results against the performance of backend models on MIR tasks. Specifically, in Table \ref{tab:mir} \textbf{Tags} column shows the AUC metric on MagnaTagATune~\cite{Law200910TI}, \textbf{Genres} shows the genre classification accuracy on GTZAN~\cite{Tzanetakis2002MusicalGC}, \textbf{Key} shows the key detection accuracy on Giantsteps~\cite{Knees2015TwoDS}, \textbf{Hot Recs} corresponds to the best variant for HitRate@50 across all recommendation models reported in this paper, and \textbf{Cold Recs} is the recommendation HitRate@20. We tried to choose the tasks and datasets reported by all the models we used. However, from the respective original papers, EncodecMAE results are not reported for tag prediction and key estimation, and MusicFM results are unavailable for genre prediction. The exact model versions we took the values for can be found in the Appendix.

We can see both commonalities and differences in the performance of backend models across MIR tasks and MRS. There is some discrepancy with MusicFM, which scores best for auto-tagging and key prediction but does not reach top scores on recommendation tasks, while still showing good performance. Notably, MusicFM is a modification of MERT, and this improvement is consistent across all tasks. Jukebox has the best results for cold recommendations, strong performance with auto-tagging and key estimation, average results for genre prediction, and the worst results for hot recommendations. MusiCNN achieves the best performance on hot recommendations but scores closer to the end on other tasks, even though its performance is comparable to auto-tagging and genre prediction. EncodecMAE shows the best results for genre prediction, comes second for \textbf{Hot recs}, and fifth for \textbf{Cold recs}. While MuQ did not achieve first place in any task, it showed stable competitive results across all the tasks. Similarly, MULE scores second in auto-tagging and key estimation, third in genre prediction, and fifth in hot recommendations; however, it underperforms in cold recommendations. MERT scores well with auto-tagging and key estimation, but slightly lower for other tasks. \mbox{Music2Vec} tends to show worse results across all MIR and MRS tasks. Poor performance of MFCC in MIR tasks translates well into Recommendation tasks, highlighting the importance of adopting more modern representations. Interestingly, MFCC performs similarly to MusiCNN in the cold test, but shows large gaps in other tasks. This stresses the fact that performance on one task does not necessarily transfer to other tasks, even within one domain: while MusiCNN is at the top and MFCC at the bottom in the hot scenario, they show similar results in the cold scenario.

\subsection{Discussion}

\subsubsection{RQ1: How do different recommendation approaches compare when initialized with pretrained audio representations?}~\\
We found that different models perform best for the hot and cold-start scenarios. For hot recommendations, BERT4Rec performed best overall, closely followed by the Hybrid model. Bimodal net and Shallow net performed less well. For cold recommendations, Bimodal net shows best performance overall as well as across different PARs.  Hybrid net performed worse than the Bimodal recommender, however with MusicFM it shows very similar performance. The list concludes with Shallow net and BERT4Rec which turned out to be less useful in the cold-start scenario.

\subsubsection{RQ2: How do different pretrained audio representations compare in the context of MRS?}~\\
MusiCNN shows the best performance overall for hot test with BERT4Rec and has good positions with Hybrid and Shallow nets, which suggests that the supervised auto-tagging task it was trained on contains much useful information for MRS. We can find support for this idea, for example, in \cite{choi2022propercontrastiveselfsupervisedlearning} where authors noted the connection between performance in genre prediction and recommendations. Jukebox showed the best result for the cold test with the Bimodal net and MusicFM shows very similar cold-start performance with Hybrid. MuQ and MuQ-MuLan follow closely and show top performance both in hot and cold cases, making them universally suitable for both tasks. They are followed by EncodecMAE, which is also applicable in both cases. Honorable mentions are MULE for the hot test and MERT for the cold test. \mbox{Music2Vec} shows slightly worse performance, which aligns with its relative performance in MIR tasks.

All PARs except for Jukebox beat the MFCC baseline for the hot test. Jukebox's poor performance for the hot test and good performance for the cold test might suggest that this representation is more focused on the content side, which is more advantageous for the cold-start scenario. This is indirectly confirmed by the fact that MFCC scores relatively higher in the cold test than in the hot test with some models.

% MusicFM shows inferior performance in our tests but outstanding performance in MIR, which may suggest some technical problems with the published model weights that we used.

\subsubsection{RQ3: How does the performance of pretrained audio representations in MRS correspond to performance in MIR tasks?}~\\
The performance can vary between different downstream tasks, the most notable difference being that MusiCNN is showing outstanding results in hot recommendations. The good performance of Jukebox in MIR tasks translates to the best result in cold recommendations but not in hot ones. On the other hand, MusicFM and MuQ show high results for all tasks. It is an interesting direction for future research to determine what causes some models to be more universally applicable and some to perform better in certain tasks. 

Cold recommendations are more aligned with MIR results than hot recommendations. However, it is safe to say that the performance in MIR tasks does not directly transfer to the performance in recommendations.

\subsection{Limitations}
One limitation of our work is the use of only one dataset, which may undermine the generalizability of our results. However, this type of research requires recommendation datasets containing raw audio files, which are rarely publicly available. The second limitation is the number and scope of the recommendation models studied. Our results show that model architecture plays a significant role in the amount of useful information extracted from content embeddings, underscoring the need for a broader range of models and architectures. Moreover, we studied only one way to incorporate embeddings into a recommender system by using them as frozen item embeddings with a learned transformation over them. However, it is also possible to try other approaches, like predicting collaborative embeddings using content information~\cite{Oord2013DeepCM}, using content embedding as a regularization on collaborative one~\cite{Magron2021NeuralCC}, or doing the parameter-efficient transfer learning \cite{ding2024parameterefficienttransferlearningmusic}. We also did not try to retrain the content models as part of a recommendation model from scratch for two reasons. Firstly, training some content models is very costly. We could not possibly retrain Jukebox or MuQ, which is reported to have been trained for two weeks using 32 GPUs. Secondly, there is ambiguity about where to draw the line between building on existing models and reproducing them from scratch. The EncodecMAE architecture trains MAE on top of the pretrained Encodec model. While we could attempt to retrain the MAE part, it raises the question of why not retrain the Encodec part also. We believe our use case is the adoption of pretrained representations when end-to-end training of the same models is not feasible. However, our work can be further improved by comparing end-to-end CNN models widely used in MRS with general-purpose pretrained MIR models. The approaches we tested could not beat a pure collaborative model, ELSA, in the hot-start scenario, undermining their applicability and calling for further research on the effective use of PARs in hybrid models for this scenario. Our work can also be improved by a more thorough study of the cold-start scenario. While we tested the relative performance of PARs in this case, comparing them with other approaches for cold-start recommendations would be interesting.

\section{Conclusion}

We compared 10 frozen pretrained audio representations for an MRS task across 5 ways of incorporating them into the recommendation process. We showed that different PARs are not necessarily equally suited to both the hot-start and cold-start scenarios. Furthermore, our experiment revealed different best-performing PARs for each case: MusiCNN, EncodecMAE, MuQ-MuLan, and MuQ for the hot-start and Jukebox, MusicFM, MuQ-MuLan, MERT, and MuQ for the cold-start. Comparing the performance of these representations across MRS and MIR tasks, we demonstrate that models that perform best on common MIR tasks do not always perform best in MRS. For example, MusiCNN is not state-of-the-art in MIR tasks, but showed the best result in our experiment for the hot-start scenario. Jukebox's good performance in MIR translates into the best result in the cold-start scenario, but it performs poorly in the hot-start scenario. However, MuQ and MuQ-MuLan constantly perform well on all tasks. We hope this paper helps inspire the adoption of pretrained audio representations in MRS.

\bibliographystyle{ACM-Reference-Format}
\bibliography{bibliography}

@article{Deldjoo2021ContentdrivenMR,
  title={Content-driven Music Recommendation: Evolution, State of the Art, and Challenges},
  author={Yashar Deldjoo and Markus Schedl and Peter Knees},
  journal={ArXiv},
  year={2021},
  volume={abs/2107.11803},
}

@article{Schedl2019DeepLI,
  title={Deep Learning in Music Recommendation Systems},
  author={Markus Schedl},
  journal={Frontiers Appl. Math. Stat.},
  year={2019},
  volume={5},
  pages={44},
}

@article{Moysis2023MusicDL,
  title={Music Deep Learning: Deep Learning Methods for Music Signal Processing—A Review of the State-of-the-Art},
  author={Lazaros Moysis and Lazaros Alexios Iliadis and Sotirios P. Sotiroudis and Achilles D. Boursianis and Maria S. Papadopoulou and Konstantinos-Iraklis D. Kokkinidis and Christos K. Volos and Panagiotis G. Sarigiannidis and Spiridon Nikolaidis and Sotirios K Goudos},
  journal={IEEE Access},
  year={2023},
  volume={11},
  pages={17031-17052},
}

@article{Shao2009MusicRB,
  title={Music Recommendation Based on Acoustic Features and User Access Patterns},
  author={Bo Shao and Mitsunori Ogihara and Dingding Wang and Tao Li},
  journal={IEEE Transactions on Audio, Speech, and Language Processing},
  year={2009},
  volume={17},
  pages={1602-1611},
}

@inproceedings{Yoshii2006HybridCA,
  title={Hybrid Collaborative and Content-based Music Recommendation Using Probabilistic Model with Latent User Preferences},
  author={Kazuyoshi Yoshii and Masataka Goto and Kazunori Komatani and Tetsuya Ogata and HIroshi G. Okuno},
  booktitle={International Society for Music Information Retrieval Conference},
  year={2006},
}

@article{Tommasel2022HaventIJ,
  title={Haven’t I just Listened to This?: Exploring Diversity in Music Recommendations},
  author={Antonela Tommasel and Juan Manuel Rodr{\'i}guez and Daniela Godoy},
  journal={Adjunct Proceedings of the 30th ACM Conference on User Modeling, Adaptation and Personalization},
  year={2022},
}

@inproceedings{Seyerlehner2009BrowsingMR,
  title={Browsing Music Recommendation Networks},
  author={Klaus Seyerlehner and Peter Knees and Dominik Schnitzer and Gerhard Widmer},
  booktitle={International Society for Music Information Retrieval Conference},
  year={2009},
}

@article{McFee2011LearningCS,
  title={Learning Content Similarity for Music Recommendation},
  author={Brian McFee and Luke Barrington and Gert R. G. Lanckriet},
  journal={IEEE Transactions on Audio, Speech, and Language Processing},
  year={2011},
  volume={20},
  pages={2207-2218},
}

@inproceedings{Chordia2008ExtendingCR,
  title={Extending Content-Based Recommendation: The Case of Indian Classical Music},
  author={Parag Chordia and Mark Godfrey and Alex Rae},
  booktitle={International Society for Music Information Retrieval Conference},
  year={2008},
}

@inproceedings{Yoshii2009ContinuousPA,
  title={Continuous pLSI and Smoothing Techniques for Hybrid Music Recommendation},
  author={Kazuyoshi Yoshii and Masataka Goto},
  booktitle={International Society for Music Information Retrieval Conference},
  year={2009},
}

@article{Schnitzer2012LocalAG,
  title={Local and global scaling reduce hubs in space},
  author={Dominik Schnitzer and Arthur Flexer and Markus Schedl and Gerhard Widmer},
  journal={J. Mach. Learn. Res.},
  year={2012},
  volume={13},
  pages={2871-2902},
}

@inproceedings{Pampalk2007MusicSunAN,
  title={MusicSun: A New Approach to Artist Recommendation},
  author={Elias Pampalk and Masataka Goto},
  booktitle={International Society for Music Information Retrieval Conference},
  year={2007},
}

@article{Andjelkovic2016MoodplayIM,
  title={Moodplay: Interactive Mood-based Music Discovery and Recommendation},
  author={Ivana Andjelkovic and Denis Parra and John O'Donovan},
  journal={Proceedings of the 2016 Conference on User Modeling Adaptation and Personalization},
  year={2016},
}

@inproceedings{Melchiorre2021LEMONSLE,
  title={LEMONS: Listenable Explanations for Music recOmmeNder Systems},
  author={Alessandro B. Melchiorre and Verena Haunschmid and Markus Schedl and Gerhard Widmer},
  booktitle={European Conference on Information Retrieval},
  year={2021},
}

@article{Park2022ExploitingNP,
  title={Exploiting Negative Preference in Content-based Music Recommendation with Contrastive Learning},
  author={Minju Park and Kyogu Lee},
  journal={Proceedings of the 16th ACM Conference on Recommender Systems},
  year={2022},
}

@article{Martijn2022KnowingMK,
  title={“Knowing me, knowing you”: personalized explanations for a music recommender system},
  author={Millecamp Martijn and Cristina Conati and Katrien Verbert},
  journal={User Modeling and User-Adapted Interaction},
  year={2022},
  volume={32},
  pages={215 - 252},
}

@article{Kaminskas2013LocationawareMR,
  title={Location-aware music recommendation using auto-tagging and hybrid matching},
  author={Marius Kaminskas and Francesco Ricci and Markus Schedl},
  journal={Proceedings of the 7th ACM conference on Recommender systems},
  year={2013},
}

@article{Cheng2016OnEL,
  title={On Effective Location-Aware Music Recommendation},
  author={Zhiyong Cheng and Jialie Shen},
  journal={ACM Transactions on Information Systems (TOIS)},
  year={2016},
  volume={34},
  pages={1 - 32},
}

@inproceedings{Schedl2014LocationAwareMA,
  title={Location-Aware Music Artist Recommendation},
  author={Markus Schedl and Dominik Schnitzer},
  booktitle={Conference on Multimedia Modeling},
  year={2014},
}

@article{lvarez2020MobileMR,
  title={Mobile music recommendations for runners based on location and emotions: The DJ-Running system},
  author={Pedro {\'A}lvarez and Francisco Javier Zarazaga-Soria and Sandra Baldassarri},
  journal={Pervasive Mob. Comput.},
  year={2020},
  volume={67},
  pages={101242},
}

@article{Yakura2022AnAS,
  title={An automated system recommending background music to listen to while working},
  author={Hiromu Yakura and Tomoyasu Nakano and Masataka Goto},
  journal={User Modeling and User-Adapted Interaction},
  year={2022},
  volume={32},
  pages={355 - 388},
}

@article{Rho2009SVRbasedMM,
  title={SVR-based music mood classification and context-based music recommendation},
  author={Seungmin Rho and Byeong-jun Han and Eenjun Hwang},
  journal={Proceedings of the 17th ACM international conference on Multimedia},
  year={2009},
}

@article{Bontempelli2022FlowMR,
  title={Flow Moods: Recommending Music by Moods on Deezer},
  author={Th{\'e}o Bontempelli and Benjamin Chapus and François Rigaud and Mathieu Morlon and Marin Lorant and Guillaume Salha-Galvan},
  journal={Proceedings of the 16th ACM Conference on Recommender Systems},
  year={2022},
}

@article{Chen2013UsingEC,
  title={Using emotional context from article for contextual music recommendation},
  author={Chih-Ming Chen and Ming-Feng Tsai and Jen-Yu Liu and Yi-Hsuan Yang},
  journal={Proceedings of the 21st ACM international conference on Multimedia},
  year={2013},
}

@article{Zangerle2020UserMF,
  title={User Models for Culture-Aware Music Recommendation: Fusing Acoustic and Cultural Cues},
  author={Eva Zangerle and Martin Pichl and Markus Schedl},
  journal={Trans. Int. Soc. Music. Inf. Retr.},
  year={2020},
  volume={3},
  pages={1-16},
}

@article{Pereira2019OnlineLT,
  title={Online learning to rank for sequential music recommendation},
  author={Bruno Laporais Pereira and Alberto Hideki Ueda and Gustavo Penha and Rodrygo L. T. Santos and Nivio Ziviani},
  journal={Proceedings of the 13th ACM Conference on Recommender Systems},
  year={2019},
}

@article{Chaves2022EfficientOL,
  title={Efficient Online Learning to Rank for Sequential Music Recommendation},
  author={Pedro Dalla Vecchia Chaves and Bruno Laporais Pereira and Rodrygo L. T. Santos},
  journal={Proceedings of the ACM Web Conference 2022},
  year={2022},
}

@article{Soleymani2015ContentbasedMR,
  title={Content-based music recommendation using underlying music preference structure},
  author={M. Soleymani and Anna Aljanaki and Frans Wiering and Remco C. Veltkamp},
  journal={2015 IEEE International Conference on Multimedia and Expo (ICME)},
  year={2015},
  pages={1-6},
}

@article{Chou2017ConditionalPN,
  title={Conditional preference nets for user and item cold start problems in music recommendation},
  author={Szu-Yu Chou and Li-Chia Yang and Yi-Hsuan Yang and Jyh-Shing Roger Jang},
  journal={2017 IEEE International Conference on Multimedia and Expo (ICME)},
  year={2017},
  pages={1147-1152},
}

@article{Pulis2021SiameseNN,
  title={Siamese Neural Networks for Content-based Cold-Start Music Recommendation.},
  author={Michael Pulis and Josef Bajada},
  journal={Proceedings of the 15th ACM Conference on Recommender Systems},
  year={2021},
}

@article{Chen2020LearningAE,
  title={Learning Audio Embeddings with User Listening Data for Content-Based Music Recommendation},
  author={K. Chen and Beici Liang and Xiaoshuang Ma and Minwei Gu},
  journal={ICASSP 2021 - 2021 IEEE International Conference on Acoustics, Speech and Signal Processing (ICASSP)},
  year={2020},
  pages={3015-3019},
}

@article{Nanopoulos2010MusicBoxPM,
  title={MusicBox: Personalized Music Recommendation Based on Cubic Analysis of Social Tags},
  author={Alexandros Nanopoulos and Dimitrios Rafailidis and Panagiotis Symeonidis and Yannis Manolopoulos},
  journal={IEEE Transactions on Audio, Speech, and Language Processing},
  year={2010},
  volume={18},
  pages={407-412},
}

@article{Cai2007ScalableMR,
  title={Scalable music recommendation by search},
  author={Rui Cai and Chao Zhang and Lei Zhang and Wei-Ying Ma},
  journal={Proceedings of the 15th ACM international conference on Multimedia},
  year={2007},
}

@article{Koch2017ProactiveCO,
  title={Proactive Caching of Music Videos based on Audio Features, Mood, and Genre},
  author={Christian Koch and Ganna Krupii and David Hausheer},
  journal={Proceedings of the 8th ACM on Multimedia Systems Conference},
  year={2017},
}

@inproceedings{Knees2006CombiningAS,
  title={Combining audio-based similarity with web-based data to accelerate automatic music playlist generation},
  author={Peter Knees and Tim Pohle and Markus Schedl and Gerhard Widmer},
  booktitle={Multimedia Information Retrieval},
  year={2006},
}

@article{Yoshii2008AnEH,
  title={An Efficient Hybrid Music Recommender System Using an Incrementally Trainable Probabilistic Generative Model},
  author={Kazuyoshi Yoshii and Masataka Goto and Kazuhiro Komatani and Tetsuya Ogata and HIroshi G. Okuno},
  journal={IEEE Transactions on Audio, Speech, and Language Processing},
  year={2008},
  volume={16},
  pages={435-447},
}

@inproceedings{McFee2011TheNL,
  title={The Natural Language of Playlists},
  author={Brian McFee and Gert R. G. Lanckriet},
  booktitle={International Society for Music Information Retrieval Conference},
  year={2011},
}

@inproceedings{Oord2013DeepCM,
  title={Deep content-based music recommendation},
  author={A{\"a}ron van den Oord and Sander Dieleman and Benjamin Schrauwen},
  booktitle={Neural Information Processing Systems},
  year={2013},
}

@article{Wang2014ImprovingCA,
  title={Improving Content-based and Hybrid Music Recommendation using Deep Learning},
  author={Xinxi Wang and Ye Wang},
  journal={Proceedings of the 22nd ACM international conference on Multimedia},
  year={2014},
}

@inproceedings{Liang2015ContentAwareCM,
  title={Content-Aware Collaborative Music Recommendation Using Pre-trained Neural Networks},
  author={Dawen Liang and Minshu Zhan and Daniel P. W. Ellis},
  booktitle={International Society for Music Information Retrieval Conference},
  year={2015},
}

@inproceedings{oramas2017deep,
  title={A deep multimodal approach for cold-start music recommendation},
  author={Oramas, Sergio and Nieto, Oriol and Sordo, Mohamed and Serra, Xavier},
  booktitle={Proceedings of the 2nd workshop on deep learning for recommender systems},
  pages={32--37},
  year={2017}
}

@article{Vall2019FeaturecombinationHR,
  title={Feature-combination hybrid recommender systems for automated music playlist continuation},
  author={Andreu Vall and Matthias Dorfer and Hamid Eghbalzadeh and Markus Schedl and Keki M. Burjorjee and Gerhard Widmer},
  journal={User Modeling and User-Adapted Interaction},
  year={2019},
  volume={29},
  pages={527-572},
}

@article{MartnGutirrez2020AME,
  title={A Multimodal End-to-End Deep Learning Architecture for Music Popularity Prediction},
  author={David Mart{\'i}n-Guti{\'e}rrez and Gustavo Hern{\'a}ndez Pe{\~n}aloza and Alberto Belmonte-Hern{\'a}ndez and Federico {\'A}lvarez Garc{\'i}a},
  journal={IEEE Access},
  year={2020},
  volume={8},
  pages={39361-39374},
}

@article{Gong2021ADM,
  title={A Deep Music Recommendation Method Based on Human Motion Analysis},
  author={Wenjuan Gong and Qingshuang Yu},
  journal={IEEE Access},
  year={2021},
  volume={9},
  pages={26290-26300},
}

@article{SheikhFathollahi2021MusicSM,
  title={Music similarity measurement and recommendation system using convolutional neural networks},
  author={Mohamadreza Sheikh Fathollahi and Farbod Razzazi},
  journal={International Journal of Multimedia Information Retrieval},
  year={2021},
  volume={10},
  pages={43 - 53},
}

@article{Gao2022AutomaticRO,
  title={Automatic Recommendation of Online Music Tracks Based on Deep Learning},
  author={Hong Gao},
  journal={Mathematical Problems in Engineering},
  year={2022},
}

@article{Adiyansjah2019MusicRS,
  title={Music Recommender System Based on Genre using Convolutional Recurrent Neural Networks},
  author={Adiyansjah and Alexander Agung Santoso Gunawan and Derwin Suhartono},
  journal={Procedia Computer Science},
  year={2019},
}

@article{choi2016automatic,
  title={Automatic tagging using deep convolutional neural networks},
  author={Choi, Keunwoo and Fazekas, George and Sandler, Mark},
  journal={arXiv preprint arXiv:1606.00298},
  year={2016}
}

@inproceedings{Miller2010GeoshuffleLC,
  title={Geoshuffle: Location-Aware, Content-based Music Browsing Using Self-organizing Tag Clouds},
  author={Scott Miller and Paul N. Reimer and S. Ness and George Tzanetakis},
  booktitle={International Society for Music Information Retrieval Conference},
  year={2010},
}

@article{Kim2018OneDM,
  title={One deep music representation to rule them all? A comparative analysis of different representation learning strategies},
  author={Jaehun Kim and Juli{\'a}n Urbano and Cynthia C. S. Liem and Alan Hanjalic},
  journal={Neural Computing and Applications},
  year={2018},
  volume={32},
  pages={1067 - 1093},
}

@article{Castellon2021CodifiedAL,
  title={Codified audio language modeling learns useful representations for music information retrieval},
  author={Rodrigo Castellon and Chris Donahue and Percy Liang},
  journal={ArXiv},
  year={2021},
  volume={abs/2107.05677},
}

@misc{spotify,
    author = "Spotify",
    title = "Audio Features Reference",
    year = "2024",
    url = "https://developer.spotify.com/documentation/web-api/reference/get-audio-features"
}

@misc{spotifyDeprecated,
    author = "Spotify",
    title = "Audio Features Deprecation",
    year = "2024",
    url = "https://developer.spotify.com/blog/2024-11-27-changes-to-the-web-api"
}

@misc{bitterLesson,
    author = "Rich Sutton",
    title = "The Bitter Lesson",
    year = "2019",
    url = "http://www.incompleteideas.net/IncIdeas/BitterLesson.html"
}

@inproceedings{BertinMahieux2011TheMS,
  author = {Thierry Bertin-Mahieux and Daniel P.W. Ellis and Brian Whitman and Paul Lamere},
  title = {The Million Song Dataset},
  booktitle = {{Proceedings of the 12th International Conference on Music Information
	Retrieval ({ISMIR} 2011)}},
  year = {2011}
}

@article{Spijkervet2021ContrastiveLO,
  title={Contrastive Learning of Musical Representations},
  author={Janne Spijkervet and John Ashley Burgoyne},
  journal={ArXiv},
  year={2021},
  volume={abs/2103.09410},
}

@article{Koo2022EndToEndMR,
  title={End-To-End Music Remastering System Using Self-Supervised And Adversarial Training},
  author={Junghyun Koo and Seungryeol Paik and Kyogu Lee},
  journal={ICASSP 2022 - 2022 IEEE International Conference on Acoustics, Speech and Signal Processing (ICASSP)},
  year={2022},
  pages={4608-4612},
}

@article{Dhariwal2020JukeboxAG,
  title={Jukebox: A Generative Model for Music},
  author={Prafulla Dhariwal and Heewoo Jun and Christine Payne and Jong Wook Kim and Alec Radford and Ilya Sutskever},
  journal={ArXiv},
  year={2020},
  volume={abs/2005.00341},
}

@article{Won2023AFM,
  title={A Foundation Model for Music Informatics},
  author={Minz Won and Yun-Ning Hung and Duc Le},
  journal={ArXiv},
  year={2023},
  volume={abs/2311.03318},
}

@article{Purwins2019DeepLF,
  title={Deep Learning for Audio Signal Processing},
  author={Hendrik Purwins and Bo Li and Tuomas Virtanen and Jan Schl\&\#x00FC;ter and Shuo-yiin Chang and Tara N. Sainath},
  journal={IEEE Journal of Selected Topics in Signal Processing},
  year={2019},
  volume={13},
  pages={206-219},
}

@article{Li2022MAPMusic2VecAS,
  title={MAP-Music2Vec: A Simple and Effective Baseline for Self-Supervised Music Audio Representation Learning},
  author={Yizhi Li and Ruibin Yuan and Ge Zhang and Yi Ma and Chenghua Lin and Xingran Chen and Anton Ragni and Hanzhi Yin and Zhijie Hu and Haoyu He and Emmanouil Benetos and Norbert Gyenge and Ruibo Liu and Jie Fu},
  journal={ArXiv},
  year={2022},
  volume={abs/2212.02508},
}

@article{Li2023MERTAM,
  title={MERT: Acoustic Music Understanding Model with Large-Scale Self-supervised Training},
  author={Yizhi Li and Ruibin Yuan and Ge Zhang and Yi Ma and Xingran Chen and Hanzhi Yin and Chen-Li Lin and Anton Ragni and Emmanouil Benetos and N. Gyenge and Roger B. Dannenberg and Ruibo Liu and Wenhu Chen and Gus G. Xia and Yemin Shi and Wen-Fen Huang and Yi-Ting Guo and Jie Fu},
  journal={ArXiv},
  year={2023},
  volume={abs/2306.00107},
}

@article{Pepino2023EnCodecMAELN,
  title={EnCodecMAE: Leveraging neural codecs for universal audio representation learning},
  author={Leonardo Pepino and Pablo Ernesto Riera and Luciana Ferrer},
  journal={ArXiv},
  year={2023},
  volume={abs/2309.07391},
}

@article{Pons2019musicnnPC,
  title={musicnn: Pre-trained convolutional neural networks for music audio tagging},
  author={Jordi Pons and Xavier Serra},
  journal={ArXiv},
  year={2019},
  volume={abs/1909.06654},
}

@inproceedings{Chiu2022SelfsupervisedLW,
  title={Self-supervised Learning with Random-projection Quantizer for Speech Recognition},
  author={Chung-Cheng Chiu and James Qin and Yu Zhang and Jiahui Yu and Yonghui Wu},
  booktitle={International Conference on Machine Learning},
  year={2022},
}

@article{Zhang2020PushingTL,
  title={Pushing the Limits of Semi-Supervised Learning for Automatic Speech Recognition},
  author={Yu Zhang and James Qin and Daniel S. Park and Wei Han and Chung-Cheng Chiu and Ruoming Pang and Quoc V. Le and Yonghui Wu},
  journal={ArXiv},
  year={2020},
  volume={abs/2010.10504},
}

@article{Defossez2022HighFN,
  title={High Fidelity Neural Audio Compression},
  author={Alexandre D'efossez and Jade Copet and Gabriel Synnaeve and Yossi Adi},
  journal={ArXiv},
  year={2022},
  volume={abs/2210.13438},
}

@article{Hsu2021HuBERTSS,
  title={HuBERT: Self-Supervised Speech Representation Learning by Masked Prediction of Hidden Units},
  author={Wei-Ning Hsu and Benjamin Bolte and Yao-Hung Hubert Tsai and Kushal Lakhotia and Ruslan Salakhutdinov and Abdel-rahman Mohamed},
  journal={IEEE/ACM Transactions on Audio, Speech, and Language Processing},
  year={2021},
  volume={29},
  pages={3451-3460},
}

@article{Abdul2022MelFC,
  title={Mel Frequency Cepstral Coefficient and its Applications: A Review},
  author={Zrar Kh. Abdul and Abdulbasit K. Al-Talabani},
  journal={IEEE Access},
  year={2022},
  volume={10},
  pages={122136-122158},
  doi={10.1109/ACCESS.2022.3223444}
}

@article{Sun2019BERT4RecSR,
  title={BERT4Rec: Sequential Recommendation with Bidirectional Encoder Representations from Transformer},
  author={Fei Sun and Jun Liu and Jian Wu and Changhua Pei and Xiao Lin and Wenwu Ou and Peng Jiang},
  journal={Proceedings of the 28th ACM International Conference on Information and Knowledge Management},
  year={2019},
}

@article{Magron2021NeuralCC,
  title={Neural content-aware collaborative filtering for cold-start music recommendation},
  author={Paul Magron and C'edric F'evotte},
  journal={Data Mining and Knowledge Discovery},
  year={2021},
  volume={36},
  pages={1971 - 2005},
}

@article{Lee2018DeepCE,
  title={Deep Content-User Embedding Model for Music Recommendation},
  author={Jongpil Lee and Kyungyun Lee and Jiyoung Park and Jangyeon Park and Juhan Nam},
  journal={ArXiv},
  year={2018},
  volume={abs/1807.06786},
}

@inproceedings{santana2020music4all,
  title={Music4all: A new music database and its applications},
  author={Santana, Igor Andr{\'e} Pegoraro and Pinhelli, Fabio and Donini, Juliano and Catharin, Leonardo and Mangolin, Rafael Biazus and Feltrim, Val{\'e}ria Delisandra and Domingues, Marcos Aur{\'e}lio and others},
  booktitle={2020 International Conference on Systems, Signals and Image Processing (IWSSIP)},
  pages={399--404},
  year={2020},
  organization={IEEE}
}

@article{Moscati2022Music4AllOnionA,
  title={Music4All-Onion -- A Large-Scale Multi-faceted Content-Centric Music Recommendation Dataset},
  author={Marta Moscati and Emilia Parada-Cabaleiro and Yashar Deldjoo and Eva Zangerle and Markus Schedl},
  journal={Proceedings of the 31st ACM International Conference on Information \& Knowledge Management},
  year={2022},
}

@inproceedings{Law200910TI,
  title={10 th International Society for Music Information Retrieval Conference ( ISMIR 2009 ) EVALUATION OF ALGORITHMS USING GAMES : THE CASE OF MUSIC TAGGING},
  author={Edith Law and Kris West and Michael I. Mandel and Mert Bay and J. S. Downie},
  year={2009},
}

@article{Tzanetakis2002MusicalGC,
  title={Musical genre classification of audio signals},
  author={George Tzanetakis and Perry R. Cook},
  journal={IEEE Trans. Speech Audio Process.},
  year={2002},
  volume={10},
  pages={293-302},
}

@inproceedings{Knees2015TwoDS,
  title={Two Data Sets for Tempo Estimation and Key Detection in Electronic Dance Music Annotated from User Corrections},
  author={Peter Knees and {\'A}ngel Faraldo and Perfecto Herrera and Richard Vogl and Sebastian B{\"o}ck and Florian H{\"o}rschl{\"a}ger and Mickael Le Goff},
  booktitle={International Society for Music Information Retrieval Conference},
  year={2015},
}

@article{Padmanabhan04072015,
author = {Jayashree Padmanabhan and Melvin Jose Johnson Premkumar},
title = {Machine Learning in Automatic Speech Recognition: A Survey},
journal = {IETE Technical Review},
volume = {32},
number = {4},
pages = {240--251},
year = {2015},
publisher = {Taylor \& Francis},
doi = {10.1080/02564602.2015.1010611}
}

@InProceedings{Li2011,
author="Li, Tom L. H.
and Chan, Antoni B.",
editor="Lee, Kuo-Tien
and Tsai, Wen-Hsiang
and Liao, Hong-Yuan Mark
and Chen, Tsuhan
and Hsieh, Jun-Wei
and Tseng, Chien-Cheng",
title="Genre Classification and the Invariance of MFCC Features to Key and Tempo",
booktitle="Advances in Multimedia Modeling",
year="2011",
publisher="Springer Berlin Heidelberg",
address="Berlin, Heidelberg",
pages="317--327"
}

@InProceedings{Friberg2011,
author="Friberg, Anders and Hedblad, Anton",
title="A Comparison of Perceptual Ratings and Computed Audio Features",
booktitle="8th Sound and Music Computing Conference",
year="2011",
doi="https://doi.org/10.5281/zenodo.849857"
}

@article{zhu2025muq,
      title={MuQ: Self-Supervised Music Representation Learning with Mel Residual Vector Quantization}, 
      author={Haina Zhu and Yizhi Zhou and Hangting Chen and Jianwei Yu and Ziyang Ma and Rongzhi Gu and Yi Luo and Wei Tan and Xie Chen},
      journal={arXiv preprint arXiv:2501.01108},
      year={2025}
}

@Article{app8010150,
AUTHOR = {Lee, Jongpil and Park, Jiyoung and Kim, Keunhyoung Luke and Nam, Juhan},
TITLE = {SampleCNN: End-to-End Deep Convolutional Neural Networks Using Very Small Filters for Music Classification},
JOURNAL = {Applied Sciences},
VOLUME = {8},
YEAR = {2018},
NUMBER = {1},
ARTICLE-NUMBER = {150},
URL = {https://www.mdpi.com/2076-3417/8/1/150},
ISSN = {2076-3417},
DOI = {10.3390/app8010150}
}

@inproceedings{elsa,
author = {Van\v{c}ura, Vojt\v{e}ch and Alves, Rodrigo and Kasalick\'{y}, Petr and Kord\'{\i}k, Pavel},
title = {Scalable Linear Shallow Autoencoder for Collaborative Filtering},
year = {2022},
isbn = {9781450392785},
publisher = {Association for Computing Machinery},
address = {New York, NY, USA},
url = {https://doi.org/10.1145/3523227.3551482},
doi = {10.1145/3523227.3551482},
booktitle = {Proceedings of the 16th ACM Conference on Recommender Systems},
pages = {604–609},
numpages = {6},
location = {Seattle, WA, USA},
series = {RecSys '22}
}

@inproceedings{beeformer, series={RecSys ’24},
   title={beeFormer: Bridging the Gap Between Semantic and Interaction Similarity in Recommender Systems},
   url={http://dx.doi.org/10.1145/3640457.3691707},
   DOI={10.1145/3640457.3691707},
   booktitle={18th ACM Conference on Recommender Systems},
   publisher={ACM},
   author={Vančura, Vojtěch and Kordík, Pavel and Straka, Milan},
   year={2024},
   month=oct, pages={1102–1107},
   collection={RecSys ’24} }

@misc{liu2019robertarobustlyoptimizedbert,
      title={RoBERTa: A Robustly Optimized BERT Pretraining Approach}, 
      author={Yinhan Liu and Myle Ott and Naman Goyal and Jingfei Du and Mandar Joshi and Danqi Chen and Omer Levy and Mike Lewis and Luke Zettlemoyer and Veselin Stoyanov},
      year={2019},
      eprint={1907.11692},
      archivePrefix={arXiv},
      primaryClass={cs.CL},
      url={https://arxiv.org/abs/1907.11692}, 
}

@misc{yeh2022decoupledcontrastivelearning,
      title={Decoupled Contrastive Learning}, 
      author={Chun-Hsiao Yeh and Cheng-Yao Hong and Yen-Chi Hsu and Tyng-Luh Liu and Yubei Chen and Yann LeCun},
      year={2022},
      eprint={2110.06848},
      archivePrefix={arXiv},
      primaryClass={cs.LG},
      url={https://arxiv.org/abs/2110.06848}, 
}

@misc{ferraro2023contrastivelearningcrossmodalartist,
      title={Contrastive Learning for Cross-modal Artist Retrieval}, 
      author={Andres Ferraro and Jaehun Kim and Sergio Oramas and Andreas Ehmann and Fabien Gouyon},
      year={2023},
      eprint={2308.06556},
      archivePrefix={arXiv},
      primaryClass={cs.IR},
      url={https://arxiv.org/abs/2308.06556}, 
}

@inproceedings{mmsbenfricsamms24,
author = {Ganh\"{o}r, Christian and Moscati, Marta and Hausberger, Anna and Nawaz, Shah and Schedl, Markus},
title = {A Multimodal Single-Branch Embedding Network for Recommendation in Cold-Start and Missing Modality Scenarios},
year = {2024},
isbn = {9798400705052},
publisher = {Association for Computing Machinery},
address = {New York, NY, USA},
url = {https://doi.org/10.1145/3640457.3688138},
doi = {10.1145/3640457.3688138},
booktitle = {Proceedings of the 18th ACM Conference on Recommender Systems},
pages = {380–390},
numpages = {11},
location = {Bari, Italy},
series = {RecSys '24}
}

@misc{mccallum2022supervisedunsupervisedlearningaudio,
      title={Supervised and Unsupervised Learning of Audio Representations for Music Understanding}, 
      author={Matthew C. McCallum and Filip Korzeniowski and Sergio Oramas and Fabien Gouyon and Andreas F. Ehmann},
      year={2022},
      eprint={2210.03799},
      archivePrefix={arXiv},
      primaryClass={cs.SD},
      url={https://arxiv.org/abs/2210.03799}, 
}

@Inbook{Schedl2022,
author="Schedl, Markus
and Knees, Peter
and McFee, Brian
and Bogdanov, Dmitry",
title="Music Recommendation Systems: Techniques, Use Cases, and Challenges",
bookTitle="Recommender Systems Handbook",
year="2022",
publisher="Springer US",
address="New York, NY",
pages="927--971",
doi="10.1007/978-1-0716-2197-4_24"
}

@misc{choi2022propercontrastiveselfsupervisedlearning,
      title={Towards Proper Contrastive Self-supervised Learning Strategies For Music Audio Representation}, 
      author={Jeong Choi and Seongwon Jang and Hyunsouk Cho and Sehee Chung},
      year={2022},
      eprint={2207.04471},
      archivePrefix={arXiv},
      primaryClass={cs.SD},
      url={https://arxiv.org/abs/2207.04471}, 
}

@misc{ding2024parameterefficienttransferlearningmusic,
      title={Parameter-Efficient Transfer Learning for Music Foundation Models}, 
      author={Yiwei Ding and Alexander Lerch},
      year={2024},
      eprint={2411.19371},
      archivePrefix={arXiv},
      primaryClass={cs.SD},
      url={https://arxiv.org/abs/2411.19371}, 
}

@misc{wang2022learninguniversalaudiorepresentations,
      title={Towards Learning Universal Audio Representations}, 
      author={Luyu Wang and Pauline Luc and Yan Wu and Adria Recasens and Lucas Smaira and Andrew Brock and Andrew Jaegle and Jean-Baptiste Alayrac and Sander Dieleman and Joao Carreira and Aaron van den Oord},
      year={2022},
      eprint={2111.12124},
      archivePrefix={arXiv},
      primaryClass={cs.SD},
      url={https://arxiv.org/abs/2111.12124}, 
}

@inproceedings{Tamm_2021, series={RecSys ’21},
   title={Quality Metrics in Recommender Systems: Do We Calculate Metrics Consistently?},
   url={http://dx.doi.org/10.1145/3460231.3478848},
   DOI={10.1145/3460231.3478848},
   booktitle={Fifteenth ACM Conference on Recommender Systems},
   publisher={ACM},
   author={Tamm, Yan-Martin and Damdinov, Rinchin and Vasilev, Alexey},
   year={2021},
   month=sep, pages={708–713},
   collection={RecSys ’21} }

@inproceedings{Tamm_2024, series={RecSys ’24},
   title={Comparative Analysis of Pretrained Audio Representations in Music Recommender Systems},
   url={http://dx.doi.org/10.1145/3640457.3688172},
   DOI={10.1145/3640457.3688172},
   booktitle={18th ACM Conference on Recommender Systems},
   publisher={ACM},
   author={Tamm, Yan-Martin and Aljanaki, Anna},
   year={2024},
   month=oct, pages={934–938},
   collection={RecSys ’24} }

@article{zhou2025does,
  title={Does Multimodality Improve Recommender Systems as Expected? A Critical Analysis and Future Directions},
  author={Zhou, Hongyu and Zhang, Yinan and Sun, Aixin and Shen, Zhiqi},
  journal={arXiv preprint arXiv:2508.05377},
  year={2025}
}

@inproceedings{spillo2025seef,
  title={See the Movie, Hear the Song, Read the Book: Extending MovieLens-1M, Last. fm-2K, and DBbook with Multimodal Data},
  author={Spillo, Giuseppe and Musacchio, Elio and Musto, Cataldo and de Gemmis, Marco and Lops, Pasquale and Semeraro, Giovanni},
  booktitle={Proceedings of the Nineteenth ACM Conference on Recommender Systems},
  pages={847--856},
  year={2025}
}

@inproceedings{Schreiber2015ImprovingGA,
  title={Improving Genre Annotations for the Million Song Dataset},
  author={Hendrik Schreiber},
  booktitle={International Society for Music Information Retrieval Conference},
  year={2015},
  url={https://api.semanticscholar.org/CorpusID:16812873}
}

@inproceedings{pembek2025let,
  title={Let It Go? Not Quite: Addressing Item Cold Start in Sequential Recommendations with Content-Based Initialization},
  author={Pembek, Anton and Fatkulin, Artem and Klenitskiy, Anton and Vasilev, Alexey},
  booktitle={Proceedings of the Nineteenth ACM Conference on Recommender Systems},
  pages={626--631},
  year={2025}
}

@article{grotschla2024towards,
  title={Towards leveraging contrastively pretrained neural audio embeddings for recommender tasks},
  author={Gr{\"o}tschla, Florian and Str{\"a}ssle, Luca and Lanzend{\"o}rfer, Luca A and Wattenhofer, Roger},
  journal={arXiv preprint arXiv:2409.09026},
  year={2024}
}

@article{rajput2023recommender,
  title={Recommender systems with generative retrieval},
  author={Rajput, Shashank and Mehta, Nikhil and Singh, Anima and Hulikal Keshavan, Raghunandan and Vu, Trung and Heldt, Lukasz and Hong, Lichan and Tay, Yi and Tran, Vinh and Samost, Jonah and others},
  journal={Advances in Neural Information Processing Systems},
  volume={36},
  pages={10299--10315},
  year={2023}
}

@inproceedings{singh2024better,
  title={Better generalization with semantic ids: A case study in ranking for recommendations},
  author={Singh, Anima and Vu, Trung and Mehta, Nikhil and Keshavan, Raghunandan and Sathiamoorthy, Maheswaran and Zheng, Yilin and Hong, Lichan and Heldt, Lukasz and Wei, Li and Tandon, Devansh and others},
  booktitle={Proceedings of the 18th ACM Conference on Recommender Systems},
  pages={1039--1044},
  year={2024}
}

@inproceedings{mei2025semantic,
  title={Semantic ids for music recommendation},
  author={Mei, M Jeffrey and Henkel, Florian and Sandberg, Samuel E and Bembom, Oliver and Ehmann, Andreas F},
  booktitle={Proceedings of the Nineteenth ACM Conference on Recommender Systems},
  pages={1070--1073},
  year={2025}
}

\clearpage
\appendix
\section{List of MRS papers using content information}
The complete list of papers used in section \ref{mrs} can be found in Table \ref{tab:features_references}. We put MFCC in its column because of the number of papers that use this representation without any other Low Features. So, in cases where a paper is referenced both in the MFCC column and the Low Features column, it means that the paper used MFCC and some other features.

\begin{table}[h!]
\centering
\caption{Papers used for Figure \ref{fig:papers} by Year}
    \begin{tabular}{|c|c|c|c|c|c|}
        \hline
        \textbf{Year} & \textbf{MFCC} & \textbf{Low Features} & \textbf{Spotify} & \textbf{Spectrogram} & \textbf{Audio} \\
        \hline\hline
        2006 & \cite{Yoshii2006HybridCA,Knees2006CombiningAS} & \cite{Yoshii2006HybridCA,Knees2006CombiningAS} & & & \\
        \hline
        2007 & \cite{Pampalk2007MusicSunAN} & \cite{Cai2007ScalableMR} & & & \\
        \hline
        2008 & \cite{Chordia2008ExtendingCR,Yoshii2008AnEH} & \cite{Chordia2008ExtendingCR,Yoshii2008AnEH} & & & \\
        \hline
        2009 & \cite{Seyerlehner2009BrowsingMR,Yoshii2009ContinuousPA,Shao2009MusicRB} & \cite{Rho2009SVRbasedMM} & & & \\
        \hline
        2010 & \cite{Nanopoulos2010MusicBoxPM} & \cite{Miller2010GeoshuffleLC} & & & \\
        \hline
        2011 & & & \cite{McFee2011TheNL} & & \\
        \hline
        2012 & \cite{McFee2011LearningCS,Schnitzer2012LocalAG} & & & & \\
        \hline
        2013 & & \cite{Kaminskas2013LocationawareMR} & \cite{Chen2013UsingEC} & \cite{Oord2013DeepCM} & \\
        \hline
        2014 & & \cite{Schedl2014LocationAwareMA} & & \cite{Wang2014ImprovingCA} & \\
        \hline
        2015 & \cite{Liang2015ContentAwareCM} & \cite{Soleymani2015ContentbasedMR} & & & \\
        \hline
        2016 & \cite{Andjelkovic2016MoodplayIM} & \cite{Cheng2016OnEL} & & & \\
        \hline
        2017 & & \cite{Koch2017ProactiveCO} & & \cite{Chou2017ConditionalPN,oramas2017deep} & \\
        \hline
        2019 & & & \cite{Pereira2019OnlineLT} & \cite{Vall2019FeaturecombinationHR,Adiyansjah2019MusicRS} & \\
        \hline
        2020 & \cite{MartnGutirrez2020AME} & & \cite{lvarez2020MobileMR,Zangerle2020UserMF,MartnGutirrez2020AME} & \cite{Chen2020LearningAE,MartnGutirrez2020AME} &  \\
        \hline
        2021 & \cite{Gong2021ADM} & \cite{SheikhFathollahi2021MusicSM} & & \cite{Melchiorre2021LEMONSLE,Pulis2021SiameseNN,SheikhFathollahi2021MusicSM} &  \\
        \hline
        2022 & & \cite{Yakura2022AnAS} & \cite{Tommasel2022HaventIJ,Martijn2022KnowingMK,Chaves2022EfficientOL} & \cite{Bontempelli2022FlowMR,Gao2022AutomaticRO} & \cite{Park2022ExploitingNP} \\
        \hline
    \end{tabular}

\label{tab:features_references}
\end{table}

\section{Model versions}
In table \ref{tab:mir} we referenced values for the performance of pretrained models for different tasks. Here we report the exact model versions we used to get the numbers which we believe correspond to the pretrained models we used.

\begin{table}[h!]
\centering
\caption{Model versions and sources for metric values}
    \begin{tabular}{|c|c|c|}
        \hline
        \textbf{Model} & \textbf{Version} & \textbf{Source} \\
        \hline\hline
        MusicFM & FM7 & \cite{Yakura2022AnAS}\\
        \hline
        Music2Vec & 5s crop & \cite{Li2022MAPMusic2VecAS}\\
        \hline
        MERT & 330M RQ-VAE & \cite{Li2023MERTAM}\\
        \hline
        EncodecMAE & mel256-ec-base\_st  & \cite{Pepino2023EnCodecMAELN}\\
        \hline
        Jukebox & (CALM) Probing JUKEBOX  & \cite{Castellon2021CodifiedAL}\\
        \hline
        MusiCNN & (Tagging) Probing MUSICNN & \cite{Castellon2021CodifiedAL}\\
        \hline
        MFCC & (No pre-training) Probing MFCC & \cite{Castellon2021CodifiedAL}\\
        \hline
        MuQ & MuQ iter & \cite{zhu2025muq}\\
        \hline
        MULE & Musicset-Sup & \cite{mccallum2022supervisedunsupervisedlearningaudio}\\
        \hline
    \end{tabular}

\label{tab:features_references}
\end{table}

\end{document}